\documentclass[journal]{IEEEtran}
\usepackage{indentfirst}
\usepackage[dvips]{graphicx}
\usepackage{amsfonts}
\usepackage{multirow}
\usepackage{amsmath,amsthm,amssymb}
\usepackage{color,changepage}
\usepackage{algorithm}
\usepackage{algorithmic}
\usepackage{bbm}
\usepackage{verbatim}
\usepackage{makecell}
\usepackage{subfigure}
\usepackage{mathrsfs}
\usepackage{arydshln}
\usepackage{cases}
\usepackage{threeparttable}
\usepackage{extarrows}
\usepackage{array}
\usepackage{bm}
\usepackage{pifont}
\usepackage[T1]{fontenc}
\usepackage{threeparttable}
\usepackage{booktabs}
\usepackage{mdwlist}
\usepackage{hyperref}
\usepackage{cite}
\usepackage{stfloats}
\allowdisplaybreaks[4]

\definecolor{newcolor}{rgb}{0.5,0,1}

\newcolumntype{I}{!{\vrule width 2pt}}

\newtheorem{Theorem}{Theorem}

\newtheorem{Definition}{Definition}

\newtheorem{Lemma}{Lemma}

\theoremstyle{remark}
\newtheorem{Remark}{Remark}
\newtheorem{Example}{Example}

\DeclareMathAlphabet{\mathpzc}{OT1}{pzc}{m}{it}

\definecolor{newcolor}{rgb}{0.5,0,1}

\begin{document}

\title{Information-Theoretically Private Matrix Multiplication From MDS-Coded Storage}
\author{Jinbao Zhu,  Songze Li and Jie Li
\thanks{

Jinbao Zhu is with the Thrust of Internet of Things, The Hong Kong University of Science and Technology (Guangzhou), Guangzhou 510006, China (e-mail: jbzhu@ust.hk).

Songze Li is with the Thrust of Internet of Things, The Hong Kong University of Science and Technology (Guangzhou), Guangzhou 510006, China, and also with the Department of Computer Science and Engineering, The Hong Kong University of Science and Technology, Hong Kong SAR, China (e-mail: songzeli@ust.hk).

Jie Li was with the Hubei Key Laboratory of Applied Mathematics, Faculty of Mathematics and Statistics, Hubei University, Wuhan 430062, China (e-mail: jieli873@gmail.com).}}

\maketitle

\begin{abstract}
We study two problems of private matrix multiplication, over a distributed computing system consisting of a master node, and multiple servers that collectively store a family of public matrices using Maximum-Distance-Separable (MDS) codes. In the first problem of Private and Secure Matrix Multiplication (PSMM) from colluding servers, the master intends to compute the product of its confidential matrix $\mathbf{A}$ with a target matrix stored on the servers, without revealing any information about $\mathbf{A}$ and the index of target matrix to some colluding servers. In the second problem of Fully Private Matrix Multiplication (FPMM) from colluding servers, the matrix $\mathbf{A}$ is also selected from another family of public matrices stored at the servers in MDS form. In this case, the indices of the two target matrices should both be kept private from colluding servers. 

We develop novel strategies for the two PSMM and FPMM problems, which simultaneously guarantee information-theoretic data/index privacy and computation correctness. 
We compare the proposed PSMM strategy with a previous PSMM strategy with a weaker privacy guarantee (non-colluding servers), and demonstrate substantial improvements over the previous strategy in terms of communication and computation overheads.
Moreover, compared with a baseline FPMM strategy that uses the idea of Private Information Retrieval (PIR) to directly retrieve the desired matrix multiplication, the proposed FPMM strategy significantly reduces storage overhead, but slightly incurs large communication and computation overheads.
\end{abstract}
\begin{IEEEkeywords}
Distributed matrix multiplication, Data and index privacy, MDS-coded storage, Colluding servers, and Polynomial secret sharing.
\end{IEEEkeywords}

\section{Introduction}\label{Introduction}
{Distributed} computing has emerged as a natural approach to overcome computation and storage barriers when performing computationally intensive tasks over a massive amount of data, 
via dispersing the computation across many distributed servers that operate in parallel. 
While distributed computing provides significant flexibility and high computation speed, it 
also raises privacy concerns about sharing raw data with external servers, as these data may contain highly sensitive information such as medical records or financial transactions. Thus, it is of vital importance to design efficient distributed computing protocols that at the same time preserve data privacy.

\begin{figure}[htbp]
\centering
    \includegraphics[width=8.7cm]{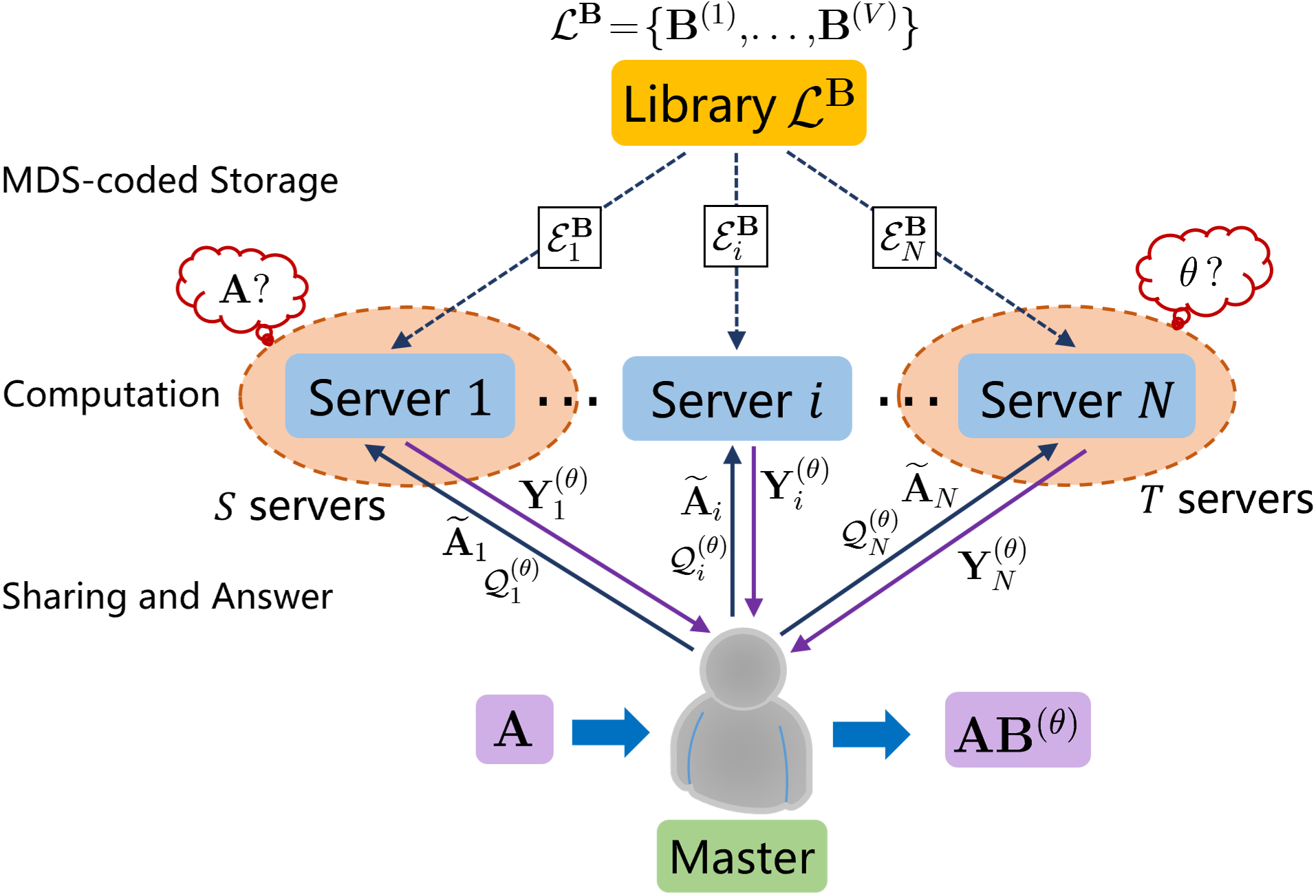} \vspace{-3mm}
    \caption{System model for private and secure matrix multiplication from MDS-coded storage with colluding servers. 
    There is a library $\mathcal{L}^{\mathbf{B}}$ that is distributedly stored across $N$ servers in an $(N, K)$ MDS-coded form. The master shares an encoding version $\widetilde{{\bf A}}_i$ of its input matrix $\mathbf{A}$ and a query ${\cal Q}_i^{(\theta)}$ with each server $i=1,2,\ldots,N$, without revealing any information about the matrix $\mathbf{A}$ and the index $\theta$ to any $S$ and $T$ colluding servers respectively. Server $i$ uses $\widetilde{{\bf A}}_i$, ${\cal Q}_i^{(\theta)}$ and its stored data $\mathcal{E}_i^{{\bf B}}$ to compute a response ${\bf Y}_i^{(\theta)}$. The master must be able to recover the product $\mathbf{A}\mathbf{B}^{(\theta)}$ from the server responses. 
    }
    \label{PSMM}
\end{figure}

\begin{figure}[htbp]
    \centering
    \includegraphics[width=8.7cm]{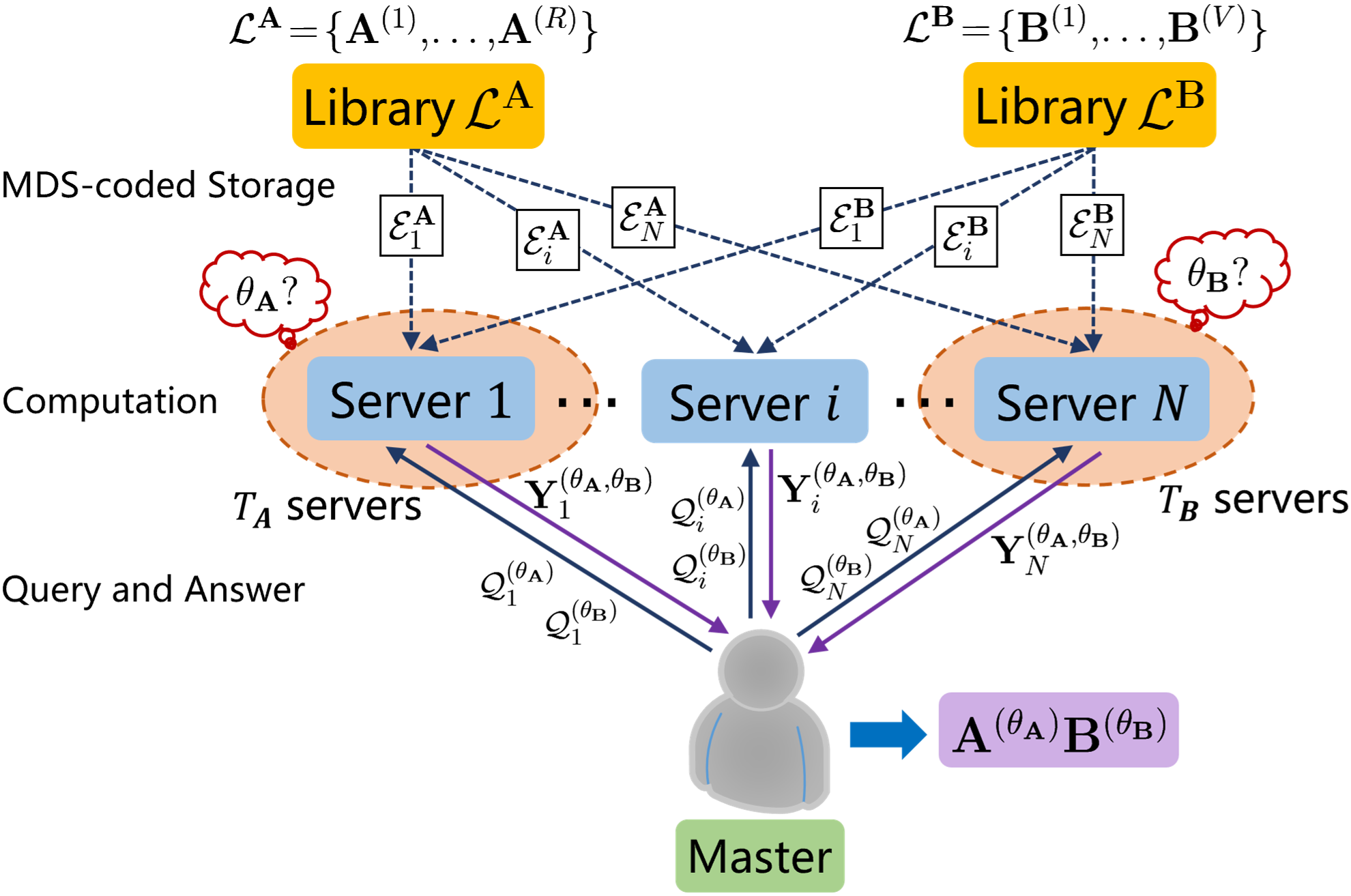} \vspace{-3mm}
    \caption{System model for fully private matrix multiplication from MDS-coded storage with colluding servers. 
    There are two libraries $\mathcal{L}^{\mathbf{A}}$ and $\mathcal{L}^{\mathbf{B}}$ that are distributedly stored across $N$ servers in $(N,K)$ MDS-coded forms. The master sends two queries ${\cal Q}_i^{(\theta_{\mathbf{A}})}$ and ${\cal Q}_i^{(\theta_{\mathbf{B}})}$ to each server $i=1,2,\ldots,N$, while keeping the indices $\theta_{\mathbf{A}}$ and $\theta_{\mathbf{B}}$ private from any $T_{\mathbf{A}}$ and $T_{\mathbf{B}}$ colluding servers respectively. Server $i$ uses ${\cal Q}_i^{(\theta_{\mathbf{A}})},{\cal Q}_i^{(\theta_{\mathbf{B}})}$ and its stored data $\mathcal{E}_i^{{\bf A}}$ and $\mathcal{E}_i^{{\bf B}}$ to compute a response ${\bf Y}_i^{(\theta_{\mathbf{A}},\theta_{\mathbf{B}})}$. The master must be able to recover the product $\mathbf{A}^{(\theta_{\mathbf{A}})}\mathbf{B}^{(\theta_{\mathbf{B}})}$  from the server responses. 
    }
    \label{FPMM}
\end{figure}

Matrix multiplication is a core building block underlying many signal processing and machine learning applications, including collaborative filtering, recommender system, and object recognition. In this paper, we focus on two \emph{private matrix multiplication} problems, where a master node would like to privately compute the product of two matrices, which are stored on a distributed file system using Maximum-Distance-Separable (MDS) codes (e.g., Reed-Solomon codes).
For the first problem of Private and Secure Matrix Multiplication (PSMM) from MDS-coded storage with colluding servers, as illustrated in Fig. \ref{PSMM}, there is a library 
of $V$ public matrices $\mathbf{B}^{(1)},\mathbf{B}^{(2)},\ldots,\mathbf{B}^{(V)}$ stored across $N$ servers following an $(N,K)$ MDS code, 
and the master owns a confidential matrix $\mathbf{A}$ and wishes to compute the product $\mathbf{A}\mathbf{B}^{(\theta)}$ 
for some $\theta=1,2,\ldots,V$, while keeping 
the matrix $\mathbf{A}$ and the index $\theta$ private from any $S$ and $T$ colluding servers respectively.
In the second problem of Fully Private Matrix Multiplication (FPMM) from MDS-coded storage with colluding servers, as shown in Fig. \ref{FPMM}, 
there is another library 
of $R$ public matrices $\mathbf{A}^{(1)},\ldots,\mathbf{A}^{(R)}$ that are also stored across the $N$ servers in MDS form. The master wants to compute $\mathbf{A}^{(\theta_{\mathbf{A}})}\mathbf{B}^{(\theta_{\mathbf{B}})}$ from the distributed system for some $\theta_{\mathbf{A}}=1,2,\ldots,R$ and $\theta_{\mathbf{B}}=1,2,\ldots,V$, while keeping the desired indices $\theta_{\mathbf{A}}$ and $\theta_{\mathbf{B}}$ private from any $T_{\mathbf{A}}$ and $T_{\mathbf{B}}$ colluding servers, where $T_{\mathbf{A}}$ and $T_{\mathbf{B}}$ are the privacy parameters for the indices $\theta_{\mathbf{A}}$ and $\theta_{\mathbf{B}}$ of  the desired computation, respectively.

As a motivating application, we consider the scenario of machine learning model inference in medical big data, where the prediction for a certain disease is achieved by computing the product of two matrices. Specifically, in a hospital, there are medical data of $R$ patients and the machine learning models for $V$ distinct diseases, e.g., diabetes, leukemia, heart disease, and so on. The data of $R$ patients are denoted by the matrices $\mathbf{A}^{(1)},\ldots,\mathbf{A}^{(R)}$ and the models for $V$ diseases are denoted by the matrices $\mathbf{B}^{(1)},\ldots,\mathbf{B}^{(V)}$. Moreover, for the sake of higher fault tolerance and storage efficiency, these data are often stored on distributed servers in a file system (locally or on the cloud), using MDS codes (see, e.g.,~\cite{ford2010availability,calder2011windows,fan2009diskreduce,dimakis2011survey}).
There is a patient (i.e., the master) who wishes to predict whether he/she suffers from some disease while keeping the identity/data of the patient private. Depending on whether the hospital has the data of the patient, the PSMM and FPMM problems are respectively considered. The FPMM problem corresponds to the case that one of the $R$ patients wishes to predict some disease, and the PSMM problem corresponds to the case that the hospital does not have the patient's medical data.
We note that while serving the patients' prediction queries, the hospital may be interested in collecting the patients' data to further improve the model performance. 
At the same time, the hospital would like to maintain a highly reliable and accurate prediction service, with an emphasis on protecting the servers from malicious attacks.
Consequently, in this paper, we consider the setting that the servers
are honest but curious,\footnote{The PSMM and FPMM strategies proposed in this paper are also robust against malicious servers who return arbitrarily erroneous results (see Remark \ref{malicious} for more detailed discussion).} which means that each server follows the protocol and correctly reports any calculations, yet a certain number of servers may be curious about the input data and potentially collude to gain information about it.

The main contribution of this paper is to develop novel strategies for the two problems of PSMM and FPMM from MDS-coded storage with colluding servers, which provide \emph{information-theoretic} privacy guarantees for the matrix $\mathbf{A}$ and the indices of desired computations, as well as correctness guarantees for the computation results. 
Our constructions are based on developing novel perfectly secure secret shares of the matrix $\mathbf{A}$ and the desired indices, which are tailored for multiplications with matrices encoded in MDS forms. Specifically, the secret shares are generated as evaluations of carefully designed polynomials, such that the computed response of each server, using the secret shares and its local Reed-Solomon (RS)-coded storage, resembles the evaluation of another composite polynomial at a particular point. Hence, having collected responses from a sufficient number of servers, the master can interpolate the polynomial, whose coefficients are designed to constitute the intended computation results.
%
%
Moreover, we analyze the communication and computation complexities of the proposed PSMM and FPMM strategies, including the upload communication cost to the servers, the download communication cost to retrieve results from the servers, the encoding and decoding complexities at the master, and the computation complexities at the servers. 

We also carry out detailed comparisons with related works in Section \ref{Comparison}.
As the most closely related work, reference~\cite{PSMM:3} considers a PSMM problem with MDS-coded storage and non-colluding servers that is a special case of our PSMM problem (i.e., $S=T=1$). While the proposed PSMM strategy provides a stronger privacy guarantee by tolerating colluding servers, it also offers substantial advantages in communication and computation efficiencies over~\cite{PSMM:3}. 
For the FPMM problem, we present a baseline strategy, where each server first computes and stores the product of each pair of encoding sub-matrices stored, and then the master  uses the idea of Private Information Retrieval (PIR) \cite{MDS-X-security} to directly retrieve the desired matrix multiplication from the results stored at servers. Compared with the baseline strategy, the proposed FPMM strategy brings a substantial advantage over storage overhead, but slightly incurs large communication and computation overheads.

\subsection{Related Work}\label{relatedwork}
Cryptography community has been investigating on addressing privacy and secrecy concerns in outsourcing the computation of matrix multiplication to an untrusted computation environment in recent years. There are some major lines of research, namely secure multiparty computation \cite{ben2019completeness,akbari2021secure,hoseini2020coded}, i.e., protocols that facilitate the computation of an arbitrary polynomial function on input matrices, homomorphic encryption \cite{jiang2018secure,duong2016efficient,rizomiliotis2022matrix} that allows performing computations on encrypted matrices without first decrypting it, verifiable matrix multiplication \cite{fiore2012publicly,liu2022privacy} that verifies whether the computation result is correct or not, and hardware-based matrix multiplication \cite{tramer2018slalom,tan2021cryptgpu,park2018efficient,hashemi2021darknight,dave2007hardware} that accelerates the execution time of matrix computation. 


These works focus on improving the computation efficiency of matrix multiplication, without leaking any information about input matrices in a computational sense with the exception of secure multiparty computation. While secure multiparty computation can provide information-theoretic privacy on input matrices in a distributed manner, it requires additional communication among all the servers for completing matrix computation.
Moreover, these works mainly put emphasis on computing two specific matrices. For the setting
of multiple matrices considered in this paper, it remains an open problem to generalize these works to provide \emph{information-theoretic} privacy protection about the indices of computation matrices. 
The PSMM and FPMM problems in this paper are different from the above works in the following two points: 1) the input matrices are selected from the libraries consisting of multiple massive matrices, and particularly the indices of the chosen matrices need to be protected in an information-theoretic sense; 2) the libraries are stored at a distributed system in the form of MDS codes, and the computation of matrix multiplication is completed over the distributed system, where the communication links among the servers are not required for completing matrix computation.
The problem of protecting the privacy of desired indices is related to the PIR problem \cite{chor1995private}. The privacy of the PIR problem can also be guaranteed in a cryptographical manner \cite{gasarch2004survey,cachin1999computationally,gentry2005single,kushilevitz1997replication} or information-theoretic manner \cite{ulukus2022private,Sun_replicated,Ulukus_MDS,X-security}. 
However, the PIR problem aims at retrieving the original files, not some functions computed from these files.
In this paper, the objective is to design information-theoretically private  computation strategies for the PSMM and FPMM problems.

Secure Matrix Multiplication (SMM) focuses on computing the product of two data matrices over a distributed computing system while keeping the data matrices secure from servers \cite{Tandon_secure_code}. 
Extensive efforts at SMM have been dedicated to using the idea of coding theory to improve the efficiency of distributed computing \cite{CIT-103,Rouayheb_secure_code,EP_SMC,Zhu_SDMM,yang2018secure,PSMM:2,Qian_Yu,batch_matrix}.
Recently, the problem of private matrix multiplication was considered in \cite{kim2019private} as an extension of SMM on the PIR setting.

\subsubsection*{Private and Secure Matrix Multiplication}
In the traditional problem of PSMM, the library of matrices is stored across \emph{non-colluding} servers in a \emph{replicated} form. In \cite{PSMM:1}, a PSMM strategy is constructed employing the ideas of  non-colluding PIR scheme  \cite{shah2014one} to create private queries and polynomial codes \cite{Polynomial_code} to complete distributed computing.  Subsequently in \cite{PSMM:Tandon}, the authors exploit the MDS-coded PIR scheme in \cite{Ulukus_MDS} to construct PSMM strategies, and characterize a tradeoff for upload and download cost, which then is improved in \cite{MDS-X-security}.
Furthermore, establishing the general tradeoff between system performance for the PSMM problem has been considered in \cite{PSMM:2,PSMM:3,Qian_Yu}. The problem of PSMM with $T$-colluding  privacy constraints was studied in \cite{PSMM:4,PSMM_Zhu}, and particularly the authors in \cite{PSMM_Zhu} present a systematic approach toward designing efficient strategies.

\subsubsection*{Fully Private Matrix Multiplication}
The problem of FPMM with replicated storage and non-colluding constraint was introduced in \cite{Qian_Yu} and a computation strategy is constructed based on Lagrange codes \cite{LCC}. Later, the problem of FPMM with $T$-colluding privacy constraint was studied in \cite{FPMM:1,PSMM_Zhu}, and the strategy in \cite{PSMM_Zhu} creates a more flexible tradeoff between system performance.

We note that the PSMM and FPMM problems considered in this paper admit a general setting for private matrix multiplication, in the sense that they include the above mentioned previous works as special cases. 
Specifically, the PSMM problem considered in this paper reduces to the PSMM problem in \cite{PSMM:1,PSMM:Tandon,PSMM:2}, by setting $K=1$ and $S=T=1$; the PSMM problem with colluding constraints in \cite{PSMM:4,PSMM_Zhu}, by setting $K=1$ and $S=T$; as well as the PSMM problem with MDS-coded storage in \cite{PSMM:3}, by setting $S=T=1$. 
The FPMM problem considered in this paper includes FPMM in \cite{Qian_Yu} and FPMM with colluding constraint in \cite{FPMM:1,PSMM_Zhu} 
 as special cases by setting $K=1,T_{\mathbf{A}}=T_{\mathbf{B}}=1$ and $K=1,T_{\mathbf{A}}=T_{\mathbf{B}}$, respectively. More additional comparisons are presented in Section \ref{Comparison}.

\subsubsection*{Notation}
The following notation is used throughout this paper.
Let boldface and cursive capital letters represent matrices and sets, respectively, e.g., $\mathbf{A}$ and $\mathcal{K}$.
For a finite set $\mathcal{K}$, $|\mathcal{K}|$ denotes its cardinality.
Denote $\mathbb{Z}^{+}$ the set of positive integers.
For nonnegative integers $m,n$ with $m<n$, $[m:n]$ and $[n]$ denote the sets $\{m,m+1,\ldots,n\}$ and $\{1,2,\ldots,n\}$, respectively.
Define $A_{\mathcal{K}}$ as $\{A_{k_1},\ldots,A_{k_{m}}\}$ for any subset $\mathcal{K}=\{k_1,\ldots,k_{m}\}\subseteq[n]$.



\section{Problem Formulations}\label{problem statement}

Consider a distributed computing system including one master node and $N$ server nodes, where all of the servers are connected to the master through error-free and orthogonal communication links.
We assume that all the servers are honest but curious,
which means that they honestly follow the prescribed protocol, yet a certain number of servers may collude to try to deduce information about private data.
We consider two private distributed matrix computation problems of \emph{private and secure matrix multiplication} and \emph{fully private matrix multiplication} from MDS-coded storage with colluding servers.
In the following, we describe the general formulations of these two problems respectively.

\subsection{Private and Secure Matrix Multiplication}
We start with the problem of Private and Secure Matrix Multiplication from MDS-coded storage with colluding servers, also referred to as PSMM problem for simplicity.
As illustrated in Fig. \ref{PSMM}, the master owns a \emph{confidential} matrix $\mathbf{A}\in\mathbb{F}^{\lambda\times\omega}$, and there is a library $\mathcal{L}^{\mathbf{B}}$ of $V$ \emph{public} matrices $\mathbf{B}^{(1)}\!,\mathbf{B}^{(2)},\!\ldots\!,\mathbf{B}^{(V)}\!\in\!\mathbb{F}^{\omega\times\gamma}$ that are distributedly stored across the $N$ servers by using a fixed $(N,K)$ MDS code over $\mathbb{F}$, for some $\lambda,\omega,\gamma\in\mathbb{Z}^{+}$ and a sufficiently large finite field $\mathbb{F}$.

Denote the encoded data of the library $\mathcal{L}^{\mathbf{B}}$ stored at server $i$ by $\mathcal{E}_{i}^{\mathbf{B}}$ for any $i\in[N]$.  The distributed system must satisfy the following storage constraint.
\begin{itemize}
  \item \textbf{MDS Property:} The library can be reconstructed by connecting to at least $K$ servers to tolerate up to $N-K$ server failures, i.e., the conditional entropy
\begin{IEEEeqnarray}{c}
H(\mathbf{B}^{(1)},\mathbf{B}^{(2)},\ldots,\mathbf{B}^{(V)}|\mathcal{E}_{\mathcal{K}}^{\mathbf{B}})=0,\;\forall\,\mathcal{K}\!\subseteq\![N],|\mathcal{K}|\geq K.\notag
\end{IEEEeqnarray}
\end{itemize}

The master privately selects an index $\theta$ from $[V]$ and wishes to compute the product $\mathbf{A}\mathbf{B}^{(\theta)}$ from the coded distributed system, while keeping its own matrix $\mathbf{A}$ secure from any colluding subset of up to $S$ out of the $N$ servers, and its desired index $\theta$ private from any $T$ colluding servers. 
For this purpose, the master employs a computation strategy 
operating in the following three phases:
\begin{itemize}
  \item \textbf{Sharing:} To provide secrecy guarantee, the master locally generates private randomness, denoted by $\mathcal{Z}^{\mathbf{A}}$, and then encodes the matrix $\mathbf{A}$ according to encoding functions $\bm{f}=(f_1,\ldots,f_{N})$, where $f_i$ corresponds to server $i$. Denote the encoded version of matrix $\mathbf{A}$ for server $i$ by $\widetilde{\mathbf{A}}_i$ which is a function of $\mathbf{A}$ and $\mathcal{Z}^{\mathbf{A}}$, i.e., $\widetilde{\mathbf{A}}_i=f_{i}(\mathbf{A},\mathcal{Z}^{\mathbf{A}})$ for all $i\in[N]$.
      
       Furthermore, to privately complete computation, the master also generates $N$ queries $\mathcal{Q}_{[N]}^{(\theta)}$, using the index $\theta$ and another  private randomness $\mathcal{Z}^{\theta}$ generated locally at the master.
       Then the encoded matrix $\widetilde{\mathbf{A}}_i$ and the query $\mathcal{Q}_{i}^{(\theta)}$ are shared with server $i\in[N]$.
  \item \textbf{Computation:}
      Upon receiving $\widetilde{\mathbf{A}}_i$ and $\mathcal{Q}_{i}^{(\theta)}$, server $i$ generates a response $\mathbf{Y}_i^{(\theta)}$, which is a deterministic function of  $\widetilde{\mathbf{A}}_i,\mathcal{Q}_{i}^{(\theta)}$ and the stored data $\mathcal{E}_{i}^{\mathbf{B}}$, i.e., $H(\mathbf{Y}_i^{(\theta)}|\widetilde{\mathbf{A}}_i,\mathcal{Q}_{i}^{(\theta)},\mathcal{E}_{i}^{\mathbf{B}})=0$ for all $i\in[N]$.
       Then server $i$ sends the response back to the master.
 \item \textbf{Reconstruction:} Due to the limits of computation and communication resources, some servers may fail to respond or respond after the master recovers the final product, who are referred to as stragglers \cite{Tail1}. For some design parameter $P\leq N$, the master only waits for the responses from the fastest $P$ servers, and then recovers the desired product $\mathbf{A}\mathbf{B}^{(\theta)}$ from their responses. This allows the computation strategy to tolerate any subset of up to $N-P$ stragglers.
\end{itemize}

The following information-theoretical constraints need to be satisfied for a valid PSMM strategy.
\begin{itemize}
\item\textbf{Privacy Constraint:} To ensure privacy constraint, the index $\theta$ of the desired product must be hidden from all the information available to any $T$ colluding servers, i.e., the mutual information
\begin{IEEEeqnarray}{c}\label{PSMM:privacy}
I(\theta;\mathcal{Q}_{\mathcal{T}}^{(\theta)},\widetilde{\mathbf{A}}_{\mathcal{T}},\mathcal{E}_{\mathcal{T}}^{\mathbf{B}},\mathbf{Y}_{\mathcal{T}}^{(\theta)})=0.
\end{IEEEeqnarray}
\item\textbf{Secrecy Constraint:} Any $S$ colluding servers must not learn any information about the  confidential matrix $\mathbf{A}$, i.e., for all $\mathcal{S}\subseteq[N],|\mathcal{S}|=S$,
\begin{IEEEeqnarray}{c}\label{cons:secu}
I(\mathbf{A};\mathcal{Q}_{\mathcal{S}}^{(\theta)},\widetilde{\mathbf{A}}_{\mathcal{S}},\mathcal{E}_{\mathcal{S}}^{\mathbf{B}},\mathbf{Y}_{\mathcal{S}}^{(\theta)})=0. 
\end{IEEEeqnarray}
\item\textbf{Correctness Constraint:} The desired product should be correctly reconstructed from the collection of responses of any fastest $P$ servers, i.e., $H(\mathbf{A}\mathbf{B}^{(\theta)}|\mathbf{Y}_{\mathcal{P}}^{(\theta)})=0$ for all $\mathcal{P}\subseteq[N],|\mathcal{P}|=P$.
\end{itemize}

The performance of a PSMM strategy can be measured using the following metrics:
\begin{enumerate}
  \item[1.] The recovery threshold $P$, which is the minimum number of servers that the master needs to wait for in order to recover the desired product $\mathbf{A}\mathbf{B}^{(\theta)}$.
  \item[2.] The communication cost,\footnote{In the current big data applications where matrix dimensions are extremely large, the upload cost for queries is negligible compared to the upload cost for matrix $\mathbf{A}$ and the download cost, as it does not scale with matrix dimensions. Similarly, in the following, the computation complexity for queries is also neglected.} which is comprised of the upload cost for matrix $\mathbf{A}$ and download cost from servers, defined as
    \begin{IEEEeqnarray}{c}
    U\!\triangleq\!\sum\limits_{i=1}^{N}H(\widetilde{\mathbf{A}}_i), \;\; 
    D\!\triangleq\! \max\limits_{\mathcal{P}:\mathcal{P}\subseteq[N],|\mathcal{P}|=P}H(\mathbf{Y}_{\mathcal{P}}^{(\theta)}). \label{upload and download} \IEEEeqnarraynumspace
    \end{IEEEeqnarray}
  \item[3.] The computation complexity, which includes the complexities of encoding, server computation, and decoding. The encoding complexity ${C}_{\mathbf{A}}$ at the master is defined as the order of the number of arithmetic operations required to compute the encoding functions $\bm{f}$. The complexity of server computation ${C}_{s}$ 
  is defined to be the order of the number of arithmetic operations required to compute the response $\mathbf{Y}_i^{(\theta)}$, maximized over $i\in[N]$. Finally, the decoding complexity ${C}_d$ at the master is defined as the order of the number of arithmetic operations required to decode the desired product $\mathbf{A}\mathbf{B}^{(\theta)}$ from the responses of the fastest servers in $\mathcal{P}$, maximized over $\mathcal{P}\subseteq[N]$ with $|\mathcal{P}|=P$.
\end{enumerate}

\subsection{Fully Private Matrix Multiplication}
We next describe the problem of Fully Private Matrix Multiplication from MDS-coded storage with colluding servers, also referred to as FPMM problem for simplicity.
Similar to the PSMM problem above, there is a library $\mathcal{L}^{\mathbf{B}}$ of $V$ public matrices $\mathbf{B}^{(1)},\mathbf{B}^{(2)},\ldots,\mathbf{B}^{(V)}\in\mathbb{F}^{\omega\times\gamma}$ distributedly stored across the $N$ servers using an $(N,K)$ MDS code over $\mathbb{F}$.
However, in the FPMM problem, as illustrated in Fig. \ref{FPMM}, 
there is another library $\mathcal{L}^{\mathbf{A}}$ of $R$ public matrices $\mathbf{A}^{(1)},\mathbf{A}^{(2)},\ldots,\mathbf{A}^{(R)}\in\mathbb{F}^{\lambda\times\omega}$ that are also stored across the $N$ servers in the form of $(N,K)$ MDS codes. 

Denote the encoded data of the libraries $\mathcal{L}^{\mathbf{A}}$ and $\mathcal{L}^{\mathbf{B}}$ stored at server $i$ by $\mathcal{E}_{i}^{\mathbf{A}}$ and $\mathcal{E}_{i}^{\mathbf{B}}$, respectively. The storage system should satisfy
\begin{itemize}
\item \textbf{Fully MDS Property:} Each of the two libraries can be reconstructed by connecting to at least $K$ servers to tolerate up to $N-K$ server failures, i.e., for all $\mathcal{K}\subseteq[N]$ with $|\mathcal{K}|\geq K$, $H(\mathbf{A}^{(1)},\ldots,\mathbf{A}^{(R)}|\mathcal{E}_{\mathcal{K}}^{\mathbf{A}})=0$ and
$H(\mathbf{B}^{(1)},\ldots,\mathbf{B}^{(V)}|\mathcal{E}_{\mathcal{K}}^{\mathbf{B}})=0$.
\end{itemize}

The master wants to compute the product $\mathbf{A}^{(\theta_{\mathbf{A}})}\mathbf{B}^{(\theta_{\mathbf{B}})}$ by utilizing the distributed computing system, while keeping the index $\theta_{\mathbf{A}}$ (resp. $\theta_{\mathbf{B}}$) private from any $T_{\mathbf{A}}$ (resp. $T_{\mathbf{B}}$) colluding servers from the information-theoretical perspective, for some $\theta_{\mathbf{A}}\in[R]$ and $\theta_{\mathbf{B}}\in[V]$.
A computation strategy for the FPMM problem operates in the following three phases:
\begin{itemize}
  \item \textbf{Sharing:}   To ensure privacy, the master generates the queries $\mathcal{Q}_{[N]}^{(\theta_{\mathbf{A}})}$ (resp. $\mathcal{Q}_{[N]}^{(\theta_{\mathbf{B}})}$), using the desired index $\theta_{\mathbf{A}}$ (resp. $\theta_{\mathbf{B}}$) and locally generated private randomness $\mathcal{Z}^{\theta_{\mathbf{A}}}$ (resp. $\mathcal{Z}^{\theta_{\mathbf{B}}}$).
  Then the two queries $\mathcal{Q}_{i}^{(\theta_{\mathbf{A}})}$ and $\mathcal{Q}_{i}^{(\theta_{\mathbf{B}})}$ are shared with server $i\in[N]$.

  \item \textbf{Computation:} Upon receiving $\mathcal{Q}_{i}^{(\theta_{\mathbf{A}})}$ and $\mathcal{Q}_{i}^{(\theta_{\mathbf{B}})}$, server $i$ computes a response $\mathbf{Y}_i^{(\theta_{\mathbf{A}},\theta_{\mathbf{B}})}$ for the master according to the received queries $\mathcal{Q}_{i}^{(\theta_{\mathbf{A}})},\mathcal{Q}_{i}^{(\theta_{\mathbf{B}})}$ and the stored data $\mathcal{E}_{i}^{\mathbf{A}},\mathcal{E}_{i}^{\mathbf{B}}$, i.e., $H(\mathbf{Y}_i^{(\theta_{\mathbf{A}},\theta_{\mathbf{B}})}|\mathcal{Q}_{i}^{(\theta_{\mathbf{A}})},\mathcal{Q}_{i}^{(\theta_{\mathbf{B}})},\mathcal{E}_{i}^{\mathbf{A}},\mathcal{E}_{i}^{\mathbf{B}})=0$ for all $i\in[N]$.
 \item \textbf{Reconstruction:} The master recovers the desired product $\mathbf{A}^{(\theta_{\mathbf{A}})}\mathbf{B}^{(\theta_{\mathbf{B}})}$ from the responses of any fastest $P$ servers.
\end{itemize}

A valid FPMM strategy must satisfy the following information-theoretical constraints.
\begin{itemize}
\item\textbf{Fully Privacy Constraint:} The desired indices $\theta_{\mathbf{A}}$ and $\theta_{\mathbf{B}}$ must be hidden from all the information available to any $T_\mathbf{A}$ and $T_{\mathbf{B}}$ colluding servers, respectively, i.e., for all $\mathcal{T}_{\mathbf{A}}\subseteq[N]$  with $|\mathcal{T}_{\mathbf{A}}|=T_{\mathbf{A}}$ and $\mathcal{T}_{\mathbf{B}}\subseteq[N]$ with  $|\mathcal{T}_{\mathbf{B}}|=T_{\mathbf{B}}$, the mutual information terms\footnote{Notably, for the FPMM problem considered in this paper, the privacy parameters $T_{\mathbf{A}}$ and $T_{\mathbf{B}}$ are allowed to be unequal. We consider the privacy constraints \eqref{security:AAA}-\eqref{security:BBB} on $\theta_{\mathbf{A}}$ and $\theta_{\mathbf{B}}$ separately for convenience.}
\begin{IEEEeqnarray}{rCl}
I(\theta_{\mathbf{A}};\mathcal{Q}_{\mathcal{T}_{\mathbf{A}}}^{(\theta_{\mathbf{A}})},\mathcal{Q}_{\mathcal{T}_{\mathbf{A}}}^{(\theta_{\mathbf{B}})},\mathcal{E}_{\mathcal{T}_{\mathbf{A}}}^{\mathbf{A}},\mathcal{E}_{\mathcal{T}_{\mathbf{A}}}^{\mathbf{B}},\mathbf{Y}_{\mathcal{T}_{\mathbf{A}}}^{(\theta_{\mathbf{A}},\theta_{\mathbf{B}})})&=&0,\label{security:AAA} \\
I(\theta_{\mathbf{B}};\mathcal{Q}_{\mathcal{T}_{\mathbf{B}}}^{(\theta_{\mathbf{A}})},\mathcal{Q}_{\mathcal{T}_{\mathbf{B}}}^{(\theta_{\mathbf{B}})},\mathcal{E}_{\mathcal{T}_{\mathbf{B}}}^{\mathbf{A}},\mathcal{E}_{\mathcal{T}_{\mathbf{B}}}^{\mathbf{B}},\mathbf{Y}_{\mathcal{T}_{\mathbf{B}}}^{(\theta_{\mathbf{A}},\theta_{\mathbf{B}})})&=&0.\label{security:BBB}
\end{IEEEeqnarray}
\item\textbf{Correctness Constraint:} The desired product must be recoverable from responses of any fastest $P$ servers,
i.e., $H(\mathbf{A}^{(\theta_{\mathbf{A}})}\mathbf{B}^{(\theta_{\mathbf{B}})}|\mathbf{Y}_{\mathcal{P}}^{(\theta_{\mathbf{A}},\theta_{\mathbf{B}})})=0$ for all $\mathcal{P}\subseteq[N],|\mathcal{P}|=P$.
\end{itemize}

Similar to the PSMM problem, the performance of an FPMM strategy is evaluated by the following key quantities: 1) the recovery threshold $P$, 2) the download cost $D$, and 3) the computation complexities consisting of generating response at each server ${C}_s$ and decoding desired product at the master ${C}_d$.


\section{Computation Strategy for PSMM}\label{strategy:PSMM}
In this section, we present our construction to solve the PSMM problem, based on novel designs of matrix and query shares. We prove that our construction achieves information-theoretic secrecy for the matrix ${\bf A}$ and information-theoretic privacy for the index $\theta$. We also analyze the computation and communication overheads of the proposed strategy.
\subsection{MDS-Coded Storage for PSMM}\label{PSMM:codedStorage}
Here we explicitly describe the coded storage system for the PSMM problem, where the library $\mathcal{L}^{\mathbf{B}}$ is stored across $N$ distributed servers in an $(N,K)$ MDS-coded form.

First, we horizontally partition each matrix in the library $\mathcal{L}^{\mathbf{B}}$ into $K$ equal-sized blocks as follows.
\begin{IEEEeqnarray}{c}\label{GT1:partion222}
\mathbf{B}^{(v)}=
\left[
  \begin{array}{@{\;}c@{\;}}
    \mathbf{B}_{1}^{(v)} \\
    \vdots  \\
    \mathbf{B}_{K}^{(v)} \\
  \end{array}
\right],\quad\forall\,v\in[V].
\end{IEEEeqnarray}
Then, for each $v\in[V]$, we construct the Reed-Solomon (RS) polynomial $e^{(v)}(x)$ of  $\mathbf{B}_{1}^{(v)},\mathbf{B}_{2}^{(v)},\ldots,\mathbf{B}_{K}^{(v)}$ as
\begin{IEEEeqnarray}{c}\label{storage:PSMM}
e^{(v)}(x)=\sum\limits_{k=1}^{K}\mathbf{B}_{k}^{(v)}x^{K-k}.
\end{IEEEeqnarray}

Let $\alpha_1,\alpha_2,\ldots,\alpha_N$ be $N$ pairwise distinct non-zero elements in $\mathbb{F}$. Then the evaluations of $\{e^{(v)}(x)\}_{v\in[V]}$ at point $x=\alpha_i$ are distributedly stored at server $i$, i.e.,
\begin{IEEEeqnarray}{c}\label{PSMM:storage2}
\mathcal{E}_{i}^{\mathbf{B}}=\left\{e^{(v)}(\alpha_i):v\in[V]\right\}.
\end{IEEEeqnarray}
Apparently, $(e^{(v)}(\alpha_1),e^{(v)}(\alpha_2),\ldots,e^{(v)}(\alpha_N))$ forms an $(N,K)$ RS codeword for any $v\in[V]$, thus such storage codes satisfy $(N,K)$ MDS property.

\subsection{General Construction for PSMM}\label{Gene:PSMM}
For some design parameters $L,M\in\mathbb{Z}^{+}$, we partition matrix ${\bf A}$ into $LK$ sub-matrices, and each block ${\bf B}_k^{(v)}$ in \eqref{GT1:partion222} into $M$ sub-matrices,\footnote{We assume that the dimensions of data matrices are sufficiently large such that $L|\lambda,K|\omega,M|\gamma$.} such that
\begin{IEEEeqnarray}{c}\label{GT1:partion1}
\mathbf{A}\!=\!
\left[
  \begin{array}{@{}c@{\;\,}c@{\;\,}c@{}}
    \mathbf{A}_{1,1}  & \ldots & \mathbf{A}_{1,K} \\
    \vdots & \ddots & \vdots \\
    \mathbf{A}_{L,1}  & \ldots & \mathbf{A}_{L,K} \\
  \end{array}
\right], \;\;
\mathbf{B}^{(v)}\!=\!
\left[
  \begin{array}{@{}c@{\;\,}c@{\;\,}c@{}}
    \mathbf{B}_{1,1}^{(v)} & \ldots & \mathbf{B}_{1,M}^{(v)} \\
    \vdots  & \ddots & \vdots \\
    \mathbf{B}_{K,1}^{(v)}  & \ldots & \mathbf{B}_{K,M}^{(v)} \\
  \end{array}
\right], \IEEEeqnarraynumspace
\end{IEEEeqnarray}
for all $v\in[V]$. Here $\mathbf{A}_{\ell,k}\in\mathbb{F}^{\frac{\lambda}{L}\times\frac{\omega}{K}}$, 
and $\mathbf{B}_{k,m}^{(v)}\in\mathbb{F}^{\frac{\omega}{K}\times\frac{\gamma}{M}}$. 
Then the desired product $\mathbf{C}^{(\theta)}=\mathbf{A}\mathbf{B}^{(\theta)}$ is given by
\begin{IEEEeqnarray}{c}\label{GT1:desred}
\mathbf{C}^{(\theta)}=\mathbf{A}\mathbf{B}^{(\theta)}=
\left[
  \begin{array}{@{}ccc@{}}
    \mathbf{C}_{1,1}^{(\theta)} & \ldots & \mathbf{C}_{1,M}^{(\theta)} \\
    \vdots & \ddots & \vdots \\
    \mathbf{C}_{L,1}^{(\theta)}  & \ldots & \mathbf{C}_{L,M}^{(\theta)} \\
  \end{array}
\right]
\end{IEEEeqnarray}
with $\mathbf{C}_{\ell,m}^{(\theta)}=\sum_{k=1}^{K}\mathbf{A}_{\ell,k}\mathbf{B}_{k,m}^{(\theta)}$ for any $\ell\in[L],m\in[M]$.

Accordingly, the encoded block $e^{(v)}(\alpha_i)$ in \eqref{storage:PSMM}-\eqref{PSMM:storage2} can be viewed as a collection of $M$ encoded sub-matrices, i.e., $e^{(v)}(\alpha_i) =(e_1^{(v)}(\alpha_i),e_2^{(v)}(\alpha_i),$ $\ldots,e_M^{(v)}(\alpha_i))$, where
\begin{IEEEeqnarray}{c}\label{PSMM:stored_code}
e_m^{(v)}(x)=\sum\limits_{k=1}^{K}\mathbf{B}_{k,m}^{(v)}x^{K-k},\quad\forall\, m\in[M].
\end{IEEEeqnarray}
Thus the storage \eqref{PSMM:storage2} at server $i$ can be equivalently written as
\begin{IEEEeqnarray}{c}\label{PSMM:Storage}
\mathcal{E}_{i}^{\mathbf{B}}=\left\{e_m^{(v)}(\alpha_i):m\in[M],v\in[V]\right\}.
\end{IEEEeqnarray}

To present a general strategy for the PSMM problem, we introduce a collection of positive integers $\{b_{\ell},d_m:\ell\in[L+1],m\in[M+1]\}$, which will be used as degree parameters for the encoding functions $f(x)$ and $h(x)$ of matrices $\mathbf{A}$ and $\mathbf{B}^{(\theta)}$ respectively. Subsequently, the encoding function $f(x)$ of $\mathbf{A}$ is exploited to create secret shares of the confidential matrix $\mathbf{A}$, and the encoding function $h(x)$ of $\mathbf{B}^{(\theta)}$ is exploited to create secret shares of the private index $\theta$, such that the responses at servers can be viewed as the evaluations of the product polynomial $g(x)\triangleq f(x)\cdot h(x)$ at distinct points and the desired matrix multiplication $\mathbf{A}\mathbf{B}^{(\theta)}$ can be completed by interpolating $g(x)$ from server responses. Naturally, this requires that the coefficients of the product polynomial $g(x)$ include the desired computation for correctness guarantee, and the strategy achieves a recovery threshold of $K=\deg(g(x))+1$. In general, the group of degree parameters $\{b_{\ell},d_m:\ell\in[L+1],m\in[M+1]\}$ is associated with a PSMM strategy with recovery threshold $K=\deg(g(x))+1$ by carefully designing the encoding functions $f(x)$ and $h(x)$ of $\mathbf{A}$ and $\mathbf{B}^{(\theta)}$ and then using these encoding functions to create secret shares of the matrix $\mathbf{A}$ and the index $\theta$.

More specifically, we first use $\{b_{\ell},d_m:\ell\in[L+1],m\in[M+1]\}$ as degree parameters to construct the following encoding polynomials of matrices $\mathbf{A}$ and $\mathbf{B}^{(\theta)}$.
\begin{IEEEeqnarray}{rCl}
f(x)\!&=&\!\sum\limits_{\ell=1}^{L}\!\!\left(\sum\limits_{k=1}^{K}\mathbf{A}_{\ell,k}x^{k-1}\right)\!x^{b_{\ell}}
+\sum\limits_{t=1}^{S}\mathbf{Z}_{t}^{\mathbf{A}}x^{b_{L+1}+t-1},\label{PSMM:encodingA} \\
h(x)\!&=&\!\sum\limits_{m=1}^{M}\!\!\left(\sum\limits_{k=1}^{K}\mathbf{B}_{k,m}^{(\theta)}x^{K-k}\!\right)\!x^{d_m}
\!+\!\!\!\!\sum\limits_{t=1}^{K+T-1} \!\!\!\mathbf{Z}_{t}^{\mathbf{B}}x^{d_{M+1}+t-1}.\label{PSMM:encodingB} \IEEEeqnarraynumspace
\end{IEEEeqnarray}
Here $\mathbf{Z}_{t}^{\mathbf{A}},t\in[S]$ and $\mathbf{Z}_{t}^{\mathbf{B}},t\in[K+T-1]$ are matrices over $\mathbb{F}$ with the same dimensions as $\mathbf{A}_{\ell,k}$ and $\mathbf{B}_{k,m}^{(\theta)}$ respectively, whose forms will be specified later.
The product of the two encoding polynomials is given by
\begin{IEEEeqnarray}{c}
g(x)=\sum\limits_{r=0}^{\delta}\mathbf{G}_r x^r=f(x)\cdot h(x) \label{product poly:PSMM}
\end{IEEEeqnarray}
with a degree
\begin{IEEEeqnarray}{l}
\delta=\max\{b_{\ell}+K-1,b_{L+1}+S-1:\ell\in[L]\}\notag\\
\quad\quad\quad+\max\{d_m\!+\!K\!-\!1,d_{M+1}\!+\!K\!+\!T\!-\!2:m\in[M]\}. \notag
\end{IEEEeqnarray}

\begin{Definition}[Achievable Degree Parameters for PSMM]\label{ADP}
The degree parameters $\{b_{\ell},d_m:\ell\in[L+1],m\in[M+1]\}$  are said to be \emph{$\delta$-achievable} for the PSMM problem if the coefficients of $g(x)$ contain all sub-matrices of $\mathbf{C}^{(\theta)}$ in \eqref{GT1:desred}, i.e., $\mathbf{C}_{\ell,m}^{(\theta)}\in\{\mathbf{G}_r:r\in[0:\delta]\}$ for all $\ell\in[L],m\in[M]$. 
\end{Definition}




Next, we continue to describe the proposed PSMM strategy, using the constructed encoding polynomials with $\delta$-achievable degree parameters. First, we provide a simple example to illustrate the key ideas.


\begin{Example}
Consider the PSMM problem with parameters $V=K=L=M=S=T=2$. The matrices $\mathbf{A}$ and $\mathbf{B}^{(1)},\mathbf{B}^{(2)}$ are partitioned as follows.
\begin{IEEEeqnarray}{c}
\mathbf{A}\!\!=\!\!
\left[
  \begin{array}{@{}c@{\;}c@{}}
    \mathbf{A}_{1,1} & \mathbf{A}_{1,2} \\
    \mathbf{A}_{2,1} & \mathbf{A}_{2,2} \\
  \end{array}
\right]\!, \;\mathbf{B}^{(1)}\!\!=\!\!
\left[
  \begin{array}{@{}c@{\;}c@{}}
    \mathbf{B}_{1,1}^{(1)} & \mathbf{B}_{1,2}^{(1)} \\
    \mathbf{B}_{2,1}^{(1)} & \mathbf{B}_{2,2}^{(1)} \\
  \end{array}
\right]\!, \;
\mathbf{B}^{(2)}\!\!=\!\!
\left[
  \begin{array}{@{}c@{\;}c@{}}
    \mathbf{B}_{1,1}^{(2)} & \mathbf{B}_{1,2}^{(2)} \\
    \mathbf{B}_{2,1}^{(2)} & \mathbf{B}_{2,2}^{(2)} \\
  \end{array}
\right]\!. \notag
\end{IEEEeqnarray}

The data stored at server $i$ is given by
\begin{IEEEeqnarray}{c}
\mathcal{E}_{i}^{\mathbf{B}}=\left\{
  \begin{array}{@{}l@{}}
    e_1^{(1)}(\alpha_i)=\mathbf{B}_{2,1}^{(1)}+\mathbf{B}_{1,1}^{(1)}\alpha_i, \\ e_2^{(1)}(\alpha_i)=\mathbf{B}_{2,2}^{(1)}+\mathbf{B}_{1,2}^{(1)}\alpha_i, \\
    e_1^{(2)}(\alpha_i)=\mathbf{B}_{2,1}^{(2)}+\mathbf{B}_{1,1}^{(2)}\alpha_i, \\ e_2^{(2)}(\alpha_i)=\mathbf{B}_{2,2}^{(2)}+\mathbf{B}_{1,2}^{(2)}\alpha_i
  \end{array}
\right\}. \notag
\end{IEEEeqnarray}

Assume that the master wishes to compute $\mathbf{C}^{(1)}=\mathbf{A}\mathbf{B}^{(1)}$,
while keeping the matrix $\mathbf{A}$ secure from $S=2$ colluding servers and the intended matrix index $\theta=1$ private from $T=2$ colluding servers. 

The desired computation has the form of
\begin{IEEEeqnarray}{rCl}
\mathbf{C}^{(1)}&=&
\left[
  \begin{array}{@{}cc@{}}
    \mathbf{C}_{1,1}^{(1)} & \mathbf{C}_{1,2}^{(1)} \\
    \mathbf{C}_{2,1}^{(1)} & \mathbf{C}_{2,2}^{(1)} \\
  \end{array}
\right]\notag \\
&=&\left[
  \begin{array}{@{}cc@{}}
    \mathbf{A}_{1,1}\mathbf{B}_{1,1}^{(1)}+\mathbf{A}_{1,2}\mathbf{B}_{2,1}^{(1)} & \mathbf{A}_{1,1}\mathbf{B}_{1,2}^{(1)}+\mathbf{A}_{1,2}\mathbf{B}_{2,2}^{(1)} \\
    \mathbf{A}_{2,1}\mathbf{B}_{1,1}^{(1)}+\mathbf{A}_{2,2}\mathbf{B}_{2,1}^{(1)} & \mathbf{A}_{2,1}\mathbf{B}_{1,2}^{(1)}+\mathbf{A}_{2,2}\mathbf{B}_{2,2}^{(1)} \\
  \end{array}
\right]. \notag
\end{IEEEeqnarray}

Utilizing the degree parameters $b_{1}=0,b_{2}=7,b_{3}=11,d_{1}=0,d_{2}=2$ and $d_{3}=4$,
we construct the encoding polynomials of $\mathbf{A}$ and $\mathbf{B}^{(1)}$ as
\begin{IEEEeqnarray}{rCl}
f(x)&\!=\!&(\mathbf{A}_{1,1}\!\!+\!\!\mathbf{A}_{1,2}x)\!+\!(\mathbf{A}_{2,1}\!\!+\!\!\mathbf{A}_{2,2}x)x^{7}\!+\!\mathbf{Z}_{1}^{\mathbf{A}}x^{11}\!+\!\mathbf{Z}_{2}^{\mathbf{A}}x^{12},\notag\\
h(x)&\!=\!&(\mathbf{B}_{2,1}^{(1)}+\mathbf{B}_{1,1}^{(1)}x)+(\mathbf{B}_{2,2}^{(1)}+\mathbf{B}_{1,2}^{(1)}x)x^{2}\notag\\
&&\quad\quad\quad\quad\quad\quad\quad\quad+\mathbf{Z}_{1}^{\mathbf{B}}x^{4}+\mathbf{Z}_{2}^{\mathbf{B}}x^{5}+\mathbf{Z}_{3}^{\mathbf{B}}x^{6}, \label{example:encoding function:B} \IEEEeqnarraynumspace
\end{IEEEeqnarray}
where $\mathbf{Z}_{1}^{\mathbf{A}},\mathbf{Z}_{2}^{\mathbf{A}}$ and $\mathbf{Z}_{1}^{\mathbf{B}},\mathbf{Z}_{2}^{\mathbf{B}},\mathbf{Z}_{3}^{\mathbf{B}}$ are the matrices with corresponding dimensions that will be specified later. Here the selected degree parameters are $18$-achievable as the sub-matrices of ${\bf C}^{(1)}$ are coefficients of the 
product polynomial $g(x)=f(x)\cdot h(x)=\sum_{r=0}^{18}\mathbf{G}_r x^r$ of degree $\delta=18$, i.e., \begin{IEEEeqnarray}{rcccl}\label{eq:output}
\mathbf{G}_{1}=\mathbf{C}_{1,1}^{(1)}, \;\; \mathbf{G}_{3}=\mathbf{C}_{1,2}^{(1)}, \; \; \mathbf{G}_{8}=\mathbf{C}_{2,1}^{(1)}, \;\; \mathbf{G}_{10}=\mathbf{C}_{2,2}^{(1)}. \IEEEeqnarraynumspace
\end{IEEEeqnarray}


The master creates the encoding polynomial $f(x)$ locally, using matrix ${\bf A}$ and two masking matrices $\mathbf{Z}_{1}^{\mathbf{A}}$ and $\mathbf{Z}_{2}^{\mathbf{A}}$ whose entities are independently and uniformly sampled from $\mathbb{F}$.
Then, for each $i \in [N]$, master selects a distinct point $\alpha_i\in\mathbb{F}$, and sends a secret share $\widetilde{\mathbf{A}}_i=f(\alpha_i)$ of ${\bf A}$ to server $i$.
To privately retrieve the computation result ${\bf C}^{(1)}$, along with $\widetilde{\mathbf{A}}_i$, the master also sends the following query $\mathcal{Q}_{i}^{(1)}$ to server $i$.
\begin{IEEEeqnarray}{c}
\mathcal{Q}_{i}^{(1)}=\left\{
  \begin{array}{@{}l@{}}
    q_1^{(1)}(\alpha_i)=1+z^{(1)}_{1,1}\alpha_i^4+z^{(1)}_{1,2}\alpha_i^5, \\ q_2^{(1)}(\alpha_i)=\alpha_i^2+z^{(1)}_{2,1}\alpha_i^4+z^{(1)}_{2,2}\alpha_i^5, \\
    q_1^{(2)}(\alpha_i)=0+z^{(2)}_{1,1}\alpha_i^4+z^{(2)}_{1,2}\alpha_i^5, \\ q_2^{(2)}(\alpha_i)=0+z^{(2)}_{2,1}\alpha_i^4+z^{(2)}_{2,2}\alpha_i^5
  \end{array}
\right\}, \notag
\end{IEEEeqnarray}
where $\{z^{(1)}_{1,1},z^{(1)}_{1,2},z^{(1)}_{2,1},z^{(1)}_{2,2},z^{(2)}_{1,1},z^{(2)}_{1,2},z^{(2)}_{2,1},z^{(2)}_{2,2}\}$ are random masks independently and uniformly sampled from $\mathbb{F}$, to protect $\theta=1$ from being inferred by $T=2$ colluding servers.

Having received the query $\mathcal{Q}_{i}^{(1)}$ from the master, server $i$ encodes its stored data $\mathcal{E}_{i}^{\mathbf{B}}$ as
\begin{IEEEeqnarray*}{rCl}
\widetilde{\mathbf{B}}_i&=&q_1^{(1)}(\alpha_i)\!\cdot\! e_1^{(1)}(\alpha_i)\!+\!q_2^{(1)}(\alpha_i)\!\cdot\! e_2^{(1)}(\alpha_i)\!\notag\\
&&\quad\quad\quad\quad+q_1^{(2)}(\alpha_i)\!\cdot\!e_1^{(2)}(\alpha_i)\!+\!q_2^{(2)}(\alpha_i)\!\cdot\! e_2^{(2)}(\alpha_i) \\
&=&(\mathbf{B}_{2,1}^{(1)}\!+\!\mathbf{B}_{1,1}^{(1)}\alpha_i)\!+\!(\mathbf{B}_{2,2}^{(1)}\!+\!\mathbf{B}_{1,2}^{(1)}\alpha_i)\alpha_i^2\!\notag\\
&&\quad\quad\quad\quad\quad\quad\quad\quad+\mathbf{Z}_{1}^{\mathbf{B}}\alpha_i^{4}\!+\!\mathbf{Z}_{2}^{\mathbf{B}}\alpha_i^{5}\!+\!\mathbf{Z}_{3}^{\mathbf{B}}\alpha_i^{6}.
\end{IEEEeqnarray*}
Here  we have $\mathbf{Z}_{1}^{\mathbf{B}}=z^{(1)}_{1,1}\mathbf{B}_{2,1}^{(1)}\!+\!z^{(1)}_{2,1}\mathbf{B}_{2,2}^{(1)}\!+\!z^{(2)}_{1,1}\mathbf{B}_{2,1}^{(2)}\!+\!z^{(2)}_{2,1}\mathbf{B}_{2,2}^{(2)}$, $\mathbf{Z}_{2}^{\mathbf{B}}=z^{(1)}_{1,1}\mathbf{B}_{1,1}^{(1)}\!+\!z^{(1)}_{1,2}\mathbf{B}_{2,1}^{(1)}\!+\!z^{(1)}_{2,1}\mathbf{B}_{1,2}^{(1)}\!+\!z^{(1)}_{2,2}\mathbf{B}_{2,2}^{(1)}+z^{(2)}_{1,1}\mathbf{B}_{1,1}^{(2)}\!+\!z^{(2)}_{1,2}\mathbf{B}_{2,1}^{(2)}\!+\!z^{(2)}_{2,1}\mathbf{B}_{1,2}^{(2)}\!+\!z^{(2)}_{2,2}\mathbf{B}_{2,2}^{(2)}$, and $\mathbf{Z}_{3}^{\mathbf{B}}=z^{(1)}_{1,2}\mathbf{B}_{1,1}^{(1)}\!+\!z^{(1)}_{2,2}\mathbf{B}_{1,2}^{(1)}\!+\!z^{(2)}_{1,2}\mathbf{B}_{1,1}^{(2)}\!+\!z^{(2)}_{2,2}\mathbf{B}_{1,2}^{(2)}$, which are constant for all the servers along the dimensions corresponding to $x^4$, $x^5$, and $x^6$. We note that the $\widetilde{\mathbf{B}}_i$ is exactly the evaluation of the designed encoding polynomial $h(x)$ in \eqref{example:encoding function:B} at the point $\alpha_i$.

Next, server $i$ computes the product $\mathbf{Y}_i^{(\theta)}=\widetilde{\mathbf{A}}_i\widetilde{\mathbf{B}}_i=f(\alpha_i)\cdot h(\alpha_i)$ as a response for the master, which is equivalent to evaluating of the polynomial $g(x)=f(x)\cdot h(x)$ at point $x=\alpha_i$.
The master can interpolate $g(x)$ from the responses of any $P=\deg(g(x))+1=\delta+1=19$ servers, and then recovers the sub-matrices of ${\bf C}^{(1)}$ from the coefficients of $g(x)$ as in (\ref{eq:output}). 
This demonstrates that our PSMM strategy achieves a recovery threshold of $19$. 
\end{Example}

For the general construction of our PSMM strategy, to keep matrix $\mathbf{A}$ secure from any $S$ servers, the master samples $S$ masking matrices  $\mathbf{Z}_{1}^{\mathbf{A}},\ldots,\mathbf{Z}_{S}^{\mathbf{A}}$ in the encoding polynomial of ${\bf A}$~\eqref{PSMM:encodingA}, independently and uniformly from $\mathbb{F}^{\frac{\lambda}{L}\times\frac{\omega}{K}}$.
To keep the index $\theta$ private from any $T$ servers, the master chooses $VMT$ random masks $\{z^{(v)}_{m,t}:t\in[T],m\in[M],v\in[V]\}$ independently and uniformly from $\mathbb{F}$. Then based on the structure of the encoding polynomial in \eqref{PSMM:encodingB}, the master constructs the query polynomial $q_{m}^{(v)}(x)$ for all $m\in[M],v\in[V]$ as
\begin{IEEEeqnarray}{rCll}
q_{m}^{(v)}(x)&=&\sum\limits_{t=1}^{T}z^{(v)}_{m,t}\cdot x^{d_{M+1}+t-1}+\left\{
\begin{array}{@{}ll}
x^{d_{m}}, &\mathrm{if}\,\, v=\theta\\
0, & \mathrm{if}\,\, v\neq\theta
\end{array}
\right.. \label{query:poly} \IEEEeqnarraynumspace
\end{IEEEeqnarray}
The query polynomial $q_{m}^{(v)}(x)$ will be used as the encoding coefficient of the storage polynomial $e_m^{(v)}(x)$ to encode the data matrices stored at servers, see \eqref{encoding:B}. Specifically, when $v=\theta$, the encoding data matrix generated by the storage polynomial $e_m^{(v)}(x)$ is desired to be computed, and the coefficient $x^{d_m}$ is used to encode the desired data matrix for all $m\in[M]$. When $v\neq\theta$, the coefficient $0$ is used to eliminate the interference from the remaining undesired data matrices for all $m\in[M]$. Particularly, the masking term $\sum_{t=1}^{T}z^{(v)}_{m,t} x^{d_{M+1}+t-1}$ protects the  privacy of the desired index.

The master shares the evaluations of $f(x)$ and $\{q_{m}^{(v)}(x):m\in[M],v\in[V]\}$ at point $x=\alpha_i$ with server $i$, i.e.,
\begin{IEEEeqnarray}{rCl}
\widetilde{\mathbf{A}}_i&=&f(\alpha_i), \label{encoding:A}\\
\mathcal{Q}_{i}^{(\theta)}&=&\{q_{m}^{(v)}(\alpha_i):m\in[M],v\in[V]\}.\label{PSMM:queryB}
\end{IEEEeqnarray} 

After receiving the query $\mathcal{Q}_{i}^{(\theta)}$, server $i$ first encodes its stored data $\mathcal{E}_{i}^{\mathbf{B}}$ \eqref{PSMM:Storage} into
\begin{IEEEeqnarray}{rCl}\label{encoding:B}
\widetilde{\mathbf{B}}_i=\sum\limits_{v=1}^{V}\sum\limits_{m=1}^{M}e_m^{(v)}(\alpha_i)\cdot q_{m}^{(v)}(\alpha_i).
\end{IEEEeqnarray}

By \eqref{PSMM:stored_code}, \eqref{query:poly} and \eqref{PSMM:encodingB}, we can denote the encoding function of $\mathcal{E}_{i}^{\mathbf{B}}$ by
\begin{IEEEeqnarray}{rCl}
&&\sum\limits_{v=1}^{V}\sum\limits_{m=1}^{M}e_m^{(v)}(x)\cdot q_{m}^{(v)}(x) =\sum\limits_{m=1}^{M}\left(\sum\limits_{k=1}^{K}\mathbf{B}_{k,m}^{(\theta)}x^{K-k}\right)x^{d_{m}}\notag\\
&&+\!\sum\limits_{v=1}^{V}\!\sum\limits_{m=1}^{M}\!\!\left(\sum\limits_{k=1}^{K}\mathbf{B}_{k,m}^{(v)}x^{K-k}\!\right)\!\!\left(\sum\limits_{t=1}^{T}z^{(v)}_{m,t}\cdot x^{d_{M+1}+t-1}\!\right)\label{PSMM:expanding} \IEEEeqnarraynumspace\\
&=&\sum\limits_{m=1}^{M}\left(\sum\limits_{k=1}^{K}\mathbf{B}_{k,m}^{(\theta)}x^{K-k}\right)x^{d_{m}}+\sum\limits_{t=1}^{K+T-1}\mathbf{Z}_{t}^{\mathbf{B}}x^{d_{M+1}+t-1}\label{encoding:alignment} \IEEEeqnarraynumspace\\
&=&h(x), \label{PSMM:encodingpoly}
\end{IEEEeqnarray}
where $\mathbf{Z}_{1}^{\mathbf{B}},\ldots,\mathbf{Z}_{K+T-1}^{\mathbf{B}}$ are linear combinations of the partitioning sub-matrices $\mathbf{B}_{k,m}^{(v)},k\in[K],m\in[M],v\in[M]$ that can be obtained explicitly by expanding the second term in \eqref{PSMM:expanding}, but whose exact forms are irrelevant.
Note that $\mathbf{Z}_{1}^{\mathbf{B}},\ldots,\mathbf{Z}_{K+T-1}^{\mathbf{B}}$ are identical across all the servers. 

Each server $i$ responds to the master with the product
\begin{IEEEeqnarray}{rCl}\label{PSMM:responses}
\mathbf{Y}_i^{(\theta)}=\widetilde{\mathbf{A}}_i\widetilde{\mathbf{B}}_i,
\end{IEEEeqnarray}
which is equivalent to evaluating of the polynomial $g(x)=f(x)\cdot h(x)$ at point $x=\alpha_i$ by \eqref{product poly:PSMM}, \eqref{encoding:A}, \eqref{encoding:B} and \eqref{PSMM:encodingpoly}.
Finally, the master can interpolate $g(x)$ from the responses of any 
$P=\deg(g(x))+1=\delta+1$ 
servers. Then by the $\delta$-achievability of the selected degree parameters for the PSMM problem, the master recovers the desired computation $\mathbf{C}^{(\theta)}=\mathbf{A}\mathbf{B}^{(\theta)}$ from the coefficients of $g(x)$.

In general, for any $\delta$-achievable degree parameters, our PSMM strategy achieves a recovery threshold  of $P=\delta+1$. In other words, constructing a PSMM strategy can be done by constructing  $\delta$-achievable degree parameters. The following lemma provides a guide on how to construct achievable degree parameters.
\begin{Lemma}\label{choose:degree}
The group of degree parameters $\{b_{\ell},d_m:\ell\in[L+1],m\in[M+1]\}$ is $\delta$-achievable with $\delta=\max\{b_{\ell}+K-1,b_{L+1}+S-1:\ell\in[L]\}+\max\{d_m+K-1,d_{M+1}+K+T-2:m\in[M]\}$ for the PSMM problem if the following conditions hold.
\begin{itemize}
    \item The values $K-1+b_{\ell}+d_m$ for all $\ell\in[L],m\in[M]$ are pairwise distinct, i.e., for all $\ell,\ell'\in[L]$ and $m,m'\in[M]$ with $(\ell,m)\neq(\ell',m')$,
    \begin{IEEEeqnarray}{c}\label{Con:1}
    K-1+b_{\ell}+d_m \neq K-1+b_{\ell'}+d_{m'}.
    \end{IEEEeqnarray}
    \item For any given $\ell\in[L],m\in[M]$, 
    \begin{IEEEeqnarray}{l}
     K-1+b_{\ell}+d_m \notin \quad\quad\quad\notag\\
    \{i\!+\!b_{\ell}\!+\!d_m\!:\!\ell\!\in\![L],m\!\in\![M],i\!\in\![0\!:\!K\!-\!2]\}\notag\\
    \cup\{2K\!-\!2\!-\!i\!+\!b_{\ell}\!+\!d_m\!:\!\ell\!\in\![L],m\!\in\![M],i\!\in\![0\!:\!K\!-\!2]\}\notag\\
    \cup \{k\!+\!b_{\ell}\!+\!d_{M+1}\!+\!t\!-\!2\!:\!\ell\!\in\![L],t\!\in\![K\!+\!T\!-\!1],k\in[K]\}\notag\\
    \cup\{K\!-\!k\!+\!d_m\!+\!b_{L+1}\!+\!t\!-\!1\!:\!m\!\in\![M],t\!\in\![S],k\!\in\![K]\}\notag\\
    \cup\{b_{L+1}\!+\!d_{M+1}\!+\!t\!+\!t'\!-\!2\!:\!t\!\in\![S],t'\!\in\![K\!+\!T\!-\!1]\}.\IEEEeqnarraynumspace \label{Con:2}
    \end{IEEEeqnarray}
\end{itemize}
\end{Lemma}
\begin{IEEEproof}
From Definition \ref{ADP}, it is enough to show that the coefficients of the product polynomial $g(x)=f(x)\cdot h(x)$ in \eqref{product poly:PSMM} contain all the desired sub-matrices  $\{\mathbf{C}_{\ell,m}^{(\theta)}=\sum_{k=1}^{K}\mathbf{A}_{\ell,k}\mathbf{B}_{k,m}^{(\theta)}:\ell\in[L],m\in[M]\}$. 

By \eqref{PSMM:encodingA} and \eqref{PSMM:encodingB}, the polynomial $g(x)=f(x)\cdot h(x)$ is exactly expanded in \eqref{expanding}, where the desired terms include the desired sub-matrix $\mathbf{C}_{\ell,m}^{(\theta)}=\sum_{k=1}^{K}\mathbf{A}_{\ell,k}\mathbf{B}_{k,m}^{(\theta)}$ for all $\ell\in[L],m\in[M]\}$. Since the degree parameters satisfy the constraint in \eqref{Con:1}, there is no interference between all the desired terms in \eqref{expanding}. Moreover, the constraint in \eqref{Con:2} means that the desired terms are independent of the remaining interference terms. 
Thus the coefficients of $g(x)$ contain all the desired sub-matrices  $\{\mathbf{C}_{\ell,m}^{(\theta)}:\ell\in[L],m\in[M]\}$. 
\begin{figure*}[htbp]
\begin{IEEEeqnarray}{rCl}
g(x)&=&\underbrace{\sum\limits_{\ell=1}^{L}\sum\limits_{m=1}^{M}\left(\sum\limits_{j=0}^{K-1}\mathbf{A}_{\ell,j+1}\mathbf{B}_{j+1,m}^{(\theta)} \right)x^{K-1+b_{\ell}+d_m}}_{\text{Desired Terms}}
+\underbrace{\sum\limits_{\ell=1}^{L}\sum\limits_{m=1}^{M}\sum\limits_{i=0}^{K-2}\left(\sum\limits_{j=0}^{i}\mathbf{A}_{\ell,j+1}\mathbf{B}_{K+j-i,m}^{(\theta)} \right)x^{i+b_{\ell}+d_m}}_{\text{Interference Terms}}\notag\\
&&+\underbrace{\sum\limits_{\ell=1}^{L}\sum\limits_{m=1}^{M}\sum\limits_{i=0}^{K-2}\left(\sum\limits_{j=0}^{i}\mathbf{A}_{\ell,K-j}\mathbf{B}_{i-j+1,m}^{(\theta)} \right)x^{2K-2-i+b_{\ell}+d_m}
+\sum\limits_{\ell=1}^{L}\sum\limits_{t=1}^{K+T-1}\sum\limits_{k=1}^{K}\mathbf{A}_{\ell,k}\mathbf{Z}_{t}^{\mathbf{B}}x^{k+b_{\ell}+d_{M+1}+t-2}}_{\text{Interference Terms}}\notag\\
&&+\underbrace{\sum\limits_{m=1}^{M}\sum\limits_{t=1}^{S}\sum\limits_{k=1}^{K}\mathbf{B}_{k,m}^{(\theta)}\mathbf{Z}_{t}^{\mathbf{A}}x^{K-k+d_m+b_{L+1}+t-1}+
\sum\limits_{t=1}^{S}\sum\limits_{t'=1}^{K+T-1}\mathbf{Z}_{t}^{\mathbf{A}}\mathbf{Z}_{t'}^{\mathbf{B}}x^{b_{L+1}+d_{M+1}+t+t'-2}}_{\text{Interference Terms}}.\label{expanding}
\end{IEEEeqnarray}
\hrulefill
\end{figure*}
\end{IEEEproof}

In the following lemma, we present three explicit choices of degree parameters that have been proposed in  \cite{Zhu_SDMM,EP_SMC}. It is also easy to verify that they satisfy the achievability requirements in Lemma \ref{choose:degree}.
\begin{Lemma}\label{acheva:lemma}
The following three explicit choices of degree parameters are achievable for the PSMM problem.
\begin{itemize}
  \item $\delta_1=(L+1)(KM+K+T-1)+S-K-T-1$ by setting $b_{\ell}\!=\!(\ell-1)(KM+K+T-1),b_{L+1}\!=\!(L-1)(KM+K+T-1)+KM,d_m\!=\!(m\!-\!1)K,d_{M\!+\!1}\!\!=\!\!K\!M$ for all $\ell\!\in\![L],m\!\in\![M]$; 
  \item $\delta_2=(M+1)(LK+S)+K+T-S-3$ by setting $b_{\ell}=(\ell-1)K,b_{L+1}=LK,d_m=(m-1)(LK+S),d_{M+1}=(M-1)(LK+S)+LK$ for all $\ell\in[L],m\in[M]$; 
  \item and $\delta_3=2LKM+K+S+T-3$ by setting $b_{\ell}=(\ell-1)MK,b_{L+1}=LKM,d_m=(m-1)K,d_{M+1}=LKM$ for all $\ell\in[L],m\in[M]$. 
\end{itemize}
\end{Lemma}

Consequently, using the achievable degree parameters in Lemma \ref{acheva:lemma}, the proposed PSMM strategy achieves a recovery threshold of $P=\min\{\delta_1+1,\delta_2+1,\delta_3+1\}$.

\subsection{Secrecy and Privacy Analysis}\label{proof:strategy}
In this subsection, we provide secrecy and privacy proofs of the proposed PSMM strategy. 
\begin{Lemma}[Generalized Secret Sharing \cite{Shamir,Zhu_SDMM}]\label{security proof}
For any parameters $X,S,\kappa,\tau\in\mathbb{Z}^{+}$, let $\mathbf{W}_{1},\ldots,\mathbf{W}_{X}\in\mathbb{F}^{\kappa\times\tau}$ be $X$ secrets, and $\mathbf{Z}_{1},\ldots,\mathbf{Z}_{S}$ be $S$ random matrices with the same dimensions as the secrets and all the entries being independently and uniformly distributed on $\mathbb{F}$. Let $\alpha_1,\alpha_2,\ldots,\alpha_N$ be $N$ pairwise distinct elements in $\mathbb{F}$.
Define a function as
\begin{IEEEeqnarray}{c}
y(x)\!=\!\mathbf{W\!}_{1}u_1(x)\!+\!\ldots\!+\!\mathbf{W\!\!}_{X}u_X(x)\!+\!\mathbf{Z}_{1}v_1(x)\!+\!\ldots\!+\!\mathbf{Z}_{S}v_S(x), \notag
\end{IEEEeqnarray}
where $u_1(x),\ldots,u_X(x),v_1(x),\ldots,v_S(x)\in\mathbb{F}[x]$ are arbitrary deterministic functions of $x$.
If the matrix
\begin{IEEEeqnarray}{c}
\mathbf{V}=
\left[
  \begin{array}{@{\,}ccc@{\,}}
    v_1(\alpha_{i_1})  & \ldots & v_S(\alpha_{i_1}) \\
    \vdots & \ddots & \vdots \\
    v_1(\alpha_{i_S})  & \ldots & v_S(\alpha_{i_S}) \\
  \end{array}
\right]_{S\times S} \notag
\end{IEEEeqnarray}
is non-singular over $\mathbb{F}$ for any $\mathcal{S}=\{i_1,\ldots,i_S\}\subseteq[N]$ with $|\mathcal{S}|=S$, then the $S$ values $y(\alpha_{i_1}),\ldots,y(\alpha_{i_S})$  can not reveal any information about the secrets $\mathbf{W}_{1},\ldots,\mathbf{W}_{X}$, i.e.,
\begin{IEEEeqnarray}{c}
I(\mathbf{W}_{1},\ldots,\mathbf{W}_{X};y(\alpha_{i_1}),\ldots,y(\alpha_{i_S}))=0. \notag
\end{IEEEeqnarray}
\end{Lemma}

\begin{Theorem}
The proposed PSMM strategy provides an information-theoretical secrecy guarantee for the matrix $\mathbf{A}$ even if any group of up to $S$ servers collude, and provides an information-theoretical privacy guarantee for the index $\theta$ even if any group of up to $T$ servers collude.
\end{Theorem}
\begin{IEEEproof}
It is sufficient to show that the proposed PSMM strategy satisfies the secrecy constraint in \eqref{cons:secu} and the privacy constraint in \eqref{PSMM:privacy}.
\subsubsection*{secrecy}
For any subset $\mathcal{S}\subseteq [N]$ of size $S$, we have the mutual information
\begin{IEEEeqnarray}{rCl}
&&I(\mathbf{A};\mathcal{Q}_{\mathcal{S}}^{(\theta)},\widetilde{\mathbf{A}}_{\mathcal{S}},\mathcal{E}_{\mathcal{S}}^{\mathbf{B}},\mathbf{Y}_{\mathcal{S}}^{(\theta)})\notag\\
&\overset{(a)}{=}&\!\!I(\!\mathbf{A};\!\widetilde{\mathbf{A}}_{\mathcal{S}}\!)\!+\! I(\!\mathbf{A};\!\mathcal{Q}_{\mathcal{S}}^{(\theta)}\!,\mathcal{E}_{\mathcal{S}}^{\mathbf{B}}|\widetilde{\mathbf{A}}_{\mathcal{S}}\!)
\!+\!I(\mathbf{A};\!\mathbf{Y}_{\mathcal{S}}^{(\theta)}|\widetilde{\mathbf{A}}_{\mathcal{S}},\mathcal{Q}_{\mathcal{S}}^{(\theta)},\mathcal{E}_{\mathcal{S}}^{\mathbf{B}})\notag\\
&\overset{(b)}{=}&\!\!I(\mathbf{A};\widetilde{\mathbf{A}}_{\mathcal{S}})
\overset{(c)}{=}0. \notag
\end{IEEEeqnarray}
Here $(a)$ is by applying the chain rule of mutual information; $(b)$ is due to the fact that the queries $\mathcal{Q}_{\mathcal{S}}^{(\theta)}$ \eqref{PSMM:queryB} and the stored data $\mathcal{E}_{\mathcal{S}}^{\mathbf{B}}$ \eqref{PSMM:Storage} are generated independently of the matrix $\mathbf{A}$ and the encoded matrices  $\widetilde{\mathbf{A}}_{\mathcal{S}}$ \eqref{encoding:A} such that $0=I(\mathbf{A},\widetilde{\mathbf{A}}_{\mathcal{S}};\mathcal{Q}_{\mathcal{S}}^{(\theta)},\mathcal{E}_{\mathcal{S}}^{\mathbf{B}})\geq I(\mathbf{A};\mathcal{Q}_{\mathcal{S}}^{(\theta)},\mathcal{E}_{\mathcal{S}}^{\mathbf{B}}|\widetilde{\mathbf{A}}_{\mathcal{S}})\geq 0$, and the responses $\mathbf{Y}_{\mathcal{S}}^{(\theta)}$  are the deterministic function of $\mathcal{Q}_{\mathcal{S}}^{(\theta)},\widetilde{\mathbf{A}}_{\mathcal{S}}$ and $\mathcal{E}_{\mathcal{S}}^{\mathbf{B}}$ by \eqref{encoding:B} and \eqref{PSMM:responses} such that 
$I(\mathbf{A};\mathbf{Y}_{\mathcal{S}}^{(\theta)}|\widetilde{\mathbf{A}}_{\mathcal{S}},\mathcal{Q}_{\mathcal{S}}^{(\theta)},\mathcal{E}_{\mathcal{S}}^{\mathbf{B}})=0$;
$(c)$ follows by \eqref{encoding:A} and Lemma \ref{security proof}.

Thus the secrecy of the proposed strategy follows by \eqref{cons:secu}.

\subsubsection*{Privacy}
We next prove that the index $\theta$ of the desired computation $\mathbf{AB}^{(\theta)}$ is private for any $T$ colluding servers. For any subset $\mathcal{T}\subseteq[N]$ of size $T$, we have
\begin{IEEEeqnarray}{rCl}
&&I(\theta;\mathcal{Q}_{\mathcal{T}}^{(\theta)},\widetilde{\mathbf{A}}_{\mathcal{T}},\mathcal{E}_{\mathcal{T}}^{\mathbf{B}},\mathbf{Y}_{\mathcal{T}}^{(\theta)}) \notag\\
&=\!&I(\theta;\!\mathcal{Q}_{\mathcal{T}}^{(\theta)})\!+\!I(\theta;\!\widetilde{\mathbf{A}}_{\mathcal{T}},\!\mathcal{E}_{\mathcal{T}}^{\mathbf{B}}|\mathcal{Q}_{\mathcal{T}}^{(\theta)})
\!+\!I(\theta;\!\mathbf{Y}_{\mathcal{T}}^{(\theta)}|\mathcal{Q}_{\mathcal{T}}^{(\theta)},\!\widetilde{\mathbf{A}}_{\mathcal{T}},\!\mathcal{E}_{\mathcal{T}}^{\mathbf{B}}) \notag\\
&\overset{(a)}{=}\!&I(\theta;\mathcal{Q}_{\mathcal{T}}^{(\theta)}) \label{PSMM:proof:privacy}\\
&\overset{(b)}{=}\!&I(\theta;\{q_{m}^{(v)}(\alpha_i):i\in\mathcal{T}\}_{m\in[M],v\in[V]})
\overset{(c)}{=} 0. \notag
\end{IEEEeqnarray}
Here $(a)$ is due to the fact that the encoded data $\widetilde{\mathbf{A}}_{\mathcal{T}}$ \eqref{encoding:A} and the stored data $\mathcal{E}_{\mathcal{T}}^{\mathbf{B}}$ \eqref{PSMM:Storage} are generated independently of the queries $\mathcal{Q}_{\mathcal{T}}^{(\theta)}$ \eqref{PSMM:queryB} and the index $\theta$, and the responses $\mathbf{Y}_{\mathcal{T}}^{(\theta)}$ are the function of $\mathcal{Q}_{\mathcal{T}}^{(\theta)},\widetilde{\mathbf{A}}_{\mathcal{T}}$ and $\mathcal{E}_{\mathcal{T}}^{\mathbf{B}}$, such that $0=I(\theta,\mathcal{Q}_{\mathcal{T}}^{(\theta)};\widetilde{\mathbf{A}}_{\mathcal{T}},\mathcal{E}_{\mathcal{T}}^{\mathbf{B}})\geq I(\theta;\widetilde{\mathbf{A}}_{\mathcal{T}},\mathcal{E}_{\mathcal{T}}^{\mathbf{B}}|\mathcal{Q}_{\mathcal{T}}^{(\theta)})\geq 0$ and
$I(\theta;\mathbf{Y}_{\mathcal{T}}^{(\theta)}|\mathcal{Q}_{\mathcal{T}}^{(\theta)},\widetilde{\mathbf{A}}_{\mathcal{T}},\mathcal{E}_{\mathcal{T}}^{\mathbf{B}})=0$;
$(b)$ follows by \eqref{PSMM:queryB}; $(c)$ holds by Lemma \ref{security proof} and the fact that
 the query elements $\{q_{m}^{(v)}(\alpha_i):i\in\mathcal{T}\}$ are protected by $T$ random noises $z^{(v)}_{m,1},\ldots,z^{(v)}_{m,T}$ by \eqref{query:poly} and \eqref{PSMM:queryB}, and all the noises $z^{(v)}_{m,1},\ldots,z^{(v)}_{m,T}$ are i.i.d. uniformly distributed on $\mathbb{F}$ across all $m\in[M],v\in[V]$.

Thus, privacy follows by \eqref{PSMM:privacy}.
\end{IEEEproof}

\subsection{Complexity Analysis}\label{PSMM:CC}
In this subsection, the system performance of the proposed PSMM strategy is analyzed. 

\begin{Lemma}[Corollaries 10.8 and 10.12 in \cite{Von}]\label{com:poly}
The evaluations of a degree-$k$ polynomial at $k+1$ arbitrary points can be done in ${{O}}(k(\log k)^2\log\log k)$ arithmetic operations, and consequently, its dual problem, interpolation of a degree-$k$ polynomial from $k+1$ arbitrary points can be performed using ${{O}}(k(\log k)^2\log\log k)$ arithmetic operations.
\end{Lemma}

\begin{Theorem}\label{theorem:PSMM}
For the PSMM problem with $(N,K)$ MDS-coded storage and colluding parameters $S,T$, let $L,M$ be arbitrary partitioning parameters, then the proposed PSMM strategy can achieve
\begin{IEEEeqnarray*}{l}
\text{Recovery Threshold:} \\ 
P=\min\big\{(L+1)(KM+K+T-1)+S-K-T, \\
\quad\quad\quad\quad\;\;\; (M+1)(LK+S)+K+T-S-2, \\
\quad\quad\quad\quad\;\;\; 2LKM+K+S+T-2\big\}, \\
\text{Upload and Download Cost:}\quad (U,D)=\left(\frac{\lambda\omega N}{LK},\frac{\lambda\gamma P}{LM}\right),\\
\text{Encoding Complexity:}\;\; {C}_{\mathbf{A}}=O\left(\frac{\lambda\omega N(\log N)^2\log\log N}{LK}\right), \\
\text{Server Computation:}\quad {C}_{s}={O}\left(\frac{V\omega\gamma}{K}+\frac{\lambda\omega\gamma}{LKM}\right), \\
\text{Decoding Complexity:}\quad {C}_d={{O}}\left(\frac{\lambda\gamma P(\log P)^2\log\log P}{LM}\right).
\end{IEEEeqnarray*}
\end{Theorem}
\begin{IEEEproof}
We know from Section \ref{Gene:PSMM} that our PSMM strategy achieves the recovery threshold. 

The master uploads an encoding sub-matrix with the dimension of $\frac{\lambda}{L}\times\frac{\omega}{K}$ to each server by \eqref{encoding:A}, and downloads a matrix with the dimension of $\frac{\lambda}{L}\times\frac{\gamma}{M}$ from each of responsive servers by \eqref{PSMM:responses}.
Thus, according to \eqref{upload and download}, the strategy achieves the upload cost $U=\frac{\lambda\omega N}{LK}$ and the download cost  $D=\frac{\lambda\gamma P}{LM}$.

In terms of encoding complexity and decoding complexity, the encoding process for matrix $\mathbf{A}$ can be viewed as evaluating a polynomial of degree less than $N$ at $N$ points for $\frac{\lambda\omega}{LK}$ times by \eqref{PSMM:encodingA} and \eqref{encoding:A}, and decoding requires interpolating a polynomial of degree $P-1$ from any $P$ responses for $\frac{\lambda\gamma}{LM}$ times. By Lemma \ref{com:poly}, the encoding and decoding achieve the complexities $O(\frac{\lambda\omega N(\log N)^2\log\log N}{LK})$ and ${{O}}(\frac{\lambda\gamma P(\log P)^2\log\log P}{LM})$, respectively.
The computation at each server involves generating a linear combination of $VM$ sub-matrices with the dimension of $\frac{\omega}{K}\times\frac{\gamma}{M}$ \eqref{encoding:B}, and multiplying  two coded sub-matrices with sizes of $\frac{\lambda}{L}\times\frac{\omega}{K}$ and $\frac{\omega}{K}\times\frac{\gamma}{M}$, which require a complexity of ${O}(\frac{V\omega\gamma}{K}+\frac{\lambda\omega\gamma}{LKM})$ with straightforward matrix multiplication algorithms.
\end{IEEEproof}

\begin{Remark}\label{remark:free}
For the proposed PSMM strategy, when the system parameters $K,N,V,S,T$ are fixed, the partitioning parameters $L,M$ control the recovery threshold, the communication cost, and the computation cost. 
The optimal recovery threshold achieved by our PSMM strategy is $3K+T+S-2$ at the case of $L=M=1$, and the upload cost and encoding complexity are optimal at the case of $M=1$ and $L=\lfloor\frac{N-K-S-T+2}{2K} \rfloor$. The server computation complexity is optimal in the case that $LM$ takes the maximum value satisfying $P\leq N$. The optimal parameters for download cost can be obtained by optimizing $\arg\min_{L,M}\frac{P}{LM}$ such that $P\leq N$, similar for decoding complexity. 
In practice, one can optimize $L,M$ according to the system resources, including the number of servers, and the computation and communication capabilities, to minimize the overall computation time. For example, assume that the number of available servers is $N$, the communication speed is $s_1$ and the computation speed is $s_2$, then one can obtain the optimal parameters $L,M$  by using numerical methods to minimize the overall time as follows.
\begin{IEEEeqnarray*}{rCl}
\arg&\min_{L,M}&\; \frac{U+D}{s_1}+\frac{C_{\mathbf{A}}+C_s+C_d}{s_2},\\
&\mathrm{s.t.}&\quad\quad P\leq  N.
\end{IEEEeqnarray*}
For any given $L,M$, it is better to make the recovery threshold as low as possible. 
Here we present three groups of achievable degree parameters for achieving low recovery threshold as much as possible. How to choose the three groups of degree parameters depends on which group achieves a lower recovery threshold for any given $M,L$. The following FPMM strategy follows similar arguments.
\end{Remark}

\begin{Remark}\label{malicious}
In the PSMM strategy and the following FPMM strategy proposed in this paper, all the server responses can be viewed as evaluations of a polynomial and the master can recover its desired computation through interpolating the polynomial. Thus all the server responses form an MDS codeword. According to the property of MDS codes, the two strategies can further provide robustness against malicious servers. More specifically, if there are up to $E$ malicious servers with unknown identities who may return arbitrarily erroneous results, the master only needs to wait for responses from $2E$ additional servers in order to correctly reconstruct the final result.
\end{Remark}

\section{Computation Strategy for FPMM}\label{section:FPMM}
In this section, we present an FPMM strategy.
For the FPMM problem illustrated in Fig. \ref{FPMM}, the goal of the master is to compute the product $\mathbf{A}^{(\theta_{\mathbf{A}})}\mathbf{B}^{(\theta_{\mathbf{B}})}$ from the coded distributed computing system, without revealing any information about the indices $\theta_{\mathbf{A}}$ and $\theta_{\mathbf{B}}$ to any $T_{\mathbf{A}}$ and $T_{\mathbf{B}}$ colluding servers respectively, for any $\theta_{\mathbf{A}}\in[R]$ and $\theta_{\mathbf{B}}\in[V]$.

\subsection{MDS-Coded Storage for FPMM}
Each matrix in the library $\mathcal{L}^{\mathbf{A}}$ is vertically divided 
into $K$ equal-sized blocks, and
each matrix in the library $\mathcal{L}^{\mathbf{B}}$ is horizontally divided into $K$ equal-sized blocks, for any fixed MDS-coded parameter $K$, i.e., for all $r\in[R],v\in[V]$,
\begin{IEEEeqnarray}{c}\label{FPMM:partition}
\mathbf{A}^{(r)}=
\left[
  \begin{array}{@{}c@{\;}c@{\;}c@{}}
    \mathbf{A}_{1}^{(r)}  & \ldots & \mathbf{A}_{K}^{(r)}
  \end{array}
\right],\quad
\mathbf{B}^{(v)}=
\left[
  \begin{array}{@{}c@{}}
    \mathbf{B}_{1}^{(v)} \\
    \vdots  \\
    \mathbf{B}_{K}^{(v)} \\
  \end{array}
\right]. \IEEEeqnarraynumspace
\end{IEEEeqnarray}

The RS-coded encoding functions of $\mathbf{A}^{(r)}$ and $\mathbf{B}^{(v)}$ are constructed as
\begin{IEEEeqnarray}{rCl}
\tilde{e}^{(r)}(x)=\sum\limits_{k=1}^{K}\mathbf{A}_{k}^{(r)}x^{k-1}, \;\;  e^{(v)}(x)=\sum\limits_{k=1}^{K}\mathbf{B}_{k}^{(v)}x^{K-k}. \label{FPMM:encstore} \IEEEeqnarraynumspace
\end{IEEEeqnarray}
Then the evaluations of these encoding functions at point $x=\alpha_i$ are stored at server $i$.
\begin{IEEEeqnarray}{c}
\mathcal{E}_{i}^{\mathbf{A}}\!=\!\left\{\tilde{e}^{(r)}(\alpha_i):r\!\in\![R]\right\},\;\;
\mathcal{E}_{i}^{\mathbf{B}}\!=\!\left\{e^{(v)}(\alpha_i):v\!\in\![V]\right\}, \label{PSMM:StorageB22} \IEEEeqnarraynumspace
\end{IEEEeqnarray}
where $\alpha_1,\ldots,\alpha_N$ are $N$ non-zero evaluation points from $\mathbb{F}$.
Apparently, such storage codes satisfy $(N,K)$ MDS property.

\subsection{General Construction for FPMM}\label{proof:theorem:FPMM}
To establish the tradeoff between system performance, we divide each block of the matrices $\mathbf{A}^{(r)}$ into $L$ sub-matrices, and each block of $\mathbf{B}^{(v)}$ into $M$ sub-matrices, for some design parameters $L,M\in\mathbb{Z}^{+}$, such that for all $r\in[R],v\in[V]$,
\begin{IEEEeqnarray}{c}\label{FPMM:matrix partition}
\mathbf{A}^{(r)}\!\!=\!\!
\left[
  \begin{array}{@{}c@{\;}c@{\;}c@{}}
    \mathbf{A}_{1,1}^{(r)}  & \ldots & \mathbf{A}_{1,K}^{(r)} \\
    \vdots  & \ddots & \vdots \\
    \mathbf{A}_{L,1}^{(r)}  & \ldots & \mathbf{A}_{L,K}^{(r)} \\
  \end{array}
\right],\,\,
\mathbf{B}^{(v)}\!\!=\!\!
\left[
  \begin{array}{@{}c@{\;}c@{\;}c@{}}
    \mathbf{B}_{1,1}^{(v)} &  \ldots & \mathbf{B}_{1,M}^{(v)} \\
    \vdots & \ddots & \vdots \\
    \mathbf{B}_{K,1}^{(v)}  & \ldots & \mathbf{B}_{K,M}^{(v)} \\
  \end{array}
\right],  \IEEEeqnarraynumspace
\end{IEEEeqnarray}
where $\mathbf{A}_{\ell,k}^{(r)}\in\mathbb{F}^{\frac{\lambda}{L}\times\frac{\omega}{K}}$ and $\mathbf{B}_{k,m}^{(v)}\in\mathbb{F}^{\frac{\omega}{K}\times\frac{\gamma}{M}}$ for any $\ell\in[L],k\in[K],m\in[M]$. The desired product $\mathbf{C}^{(\theta_{\mathbf{A}},\theta_{\mathbf{B}})}=\mathbf{A}^{(\theta_{\mathbf{A}})}\mathbf{B}^{(\theta_{\mathbf{B}})}$ is given by
\begin{IEEEeqnarray}{c}\label{FPMM:desiredPro}
\mathbf{C}^{(\theta_{\mathbf{A}},\theta_{\mathbf{B}})}\!=\!\mathbf{A}^{(\theta_{\mathbf{A}})}\mathbf{B}^{(\theta_{\mathbf{B}})}\!=\!
\left[
  \begin{array}{@{\,}ccc@{\,}}
    \mathbf{C}_{1,1}^{(\theta_{\mathbf{A}},\theta_{\mathbf{B}})} & \ldots & \mathbf{C}_{1,M}^{(\theta_{\mathbf{A}},\theta_{\mathbf{B}})} \\
    \vdots & \ddots & \vdots \\
    \mathbf{C}_{L,1}^{(\theta_{\mathbf{A}},\theta_{\mathbf{B}})} & \ldots & \mathbf{C}_{L,M}^{(\theta_{\mathbf{A}},\theta_{\mathbf{B}})} \\
  \end{array}
\right] \IEEEeqnarraynumspace
\end{IEEEeqnarray}
with $\mathbf{C}_{\ell,m}^{(\theta_{\mathbf{A}},\theta_{\mathbf{B}})}\!=\!\sum_{k=1}^{K}\mathbf{A}_{\ell,k}^{(\theta_{\mathbf{A}})}\mathbf{B}_{k,m}^{(\theta_{\mathbf{B}})}$ for any $\ell\!\in\![L],m\!\in\![M]$.

Then the storage \eqref{PSMM:StorageB22} at server $i$ can be equivalently represented as
\begin{IEEEeqnarray}{rCl}
\mathcal{E}_{i}^{\mathbf{A}}&=&\left\{\tilde{e}_{\ell}^{(r)}(\alpha_i):\ell\in[L],r\in[R]\right\},\label{PSMM:StorageA}\\
\mathcal{E}_{i}^{\mathbf{B}}&=&\left\{e_m^{(v)}(\alpha_i):m\in[M],v\in[V]\right\}, \label{PSMM:StorageB}
\end{IEEEeqnarray}
where $\tilde{e}_{\ell}^{(r)}(x)$ and $e_m^{(v)}(x)$ are the encoding functions of the sub-matrices in the $\ell$-th row of the partitioning matrix $\mathbf{A}^{(r)}$ and the $m$-th column of the partitioning matrix $\mathbf{B}^{(v)}$, respectively, given by
\begin{IEEEeqnarray}{c}
\tilde{e}_{\ell}^{(r)}(x)=\sum\limits_{k=1}^{K}\mathbf{A}_{\ell,k}^{(r)}x^{k-1},\;\;\;
e_m^{(v)}(x)=\sum\limits_{k=1}^{K}\mathbf{B}_{k,m}^{(v)}x^{K-k}. \label{FPMM:stored_codeA} \IEEEeqnarraynumspace
\end{IEEEeqnarray}

Using a collection of positive integers $\{b_{\ell},d_m:\ell\in[L+1],m\in[M+1]\}$ as degree parameters for the FPMM problem, we construct the encoding polynomials of matrices $\mathbf{A}^{(\theta_{\mathbf{A}})}$ and $\mathbf{B}^{(\theta_{\mathbf{B}})}$ \eqref{FPMM:matrix partition} as
\begin{IEEEeqnarray}{rCl}
f'(x)&=&\sum\limits_{\ell=1}^{L}\left(\sum\limits_{k=1}^{K}\mathbf{A}_{\ell,k}^{(\theta_{\mathbf{A}})}x^{k-1}\right)x^{b_{\ell}}\notag\\
&&\quad\quad\quad\quad\quad\quad+\sum\limits_{t=1}^{K+T_{\mathbf{A}}-1}\mathbf{Z}_{t}^{\mathbf{A}}x^{b_{L+1}+t-1},\label{FPMM:encodingA} \IEEEeqnarraynumspace \\
h'(x)&=&\sum\limits_{m=1}^{M}\left(\sum\limits_{k=1}^{K}\mathbf{B}_{k,m}^{(\theta_{\mathbf{B}})}x^{K-k}\right)x^{d_m}\notag\\
&&\quad\quad\quad\quad\quad\quad+\sum\limits_{t=1}^{K+T_{\mathbf{B}}-1}\mathbf{Z}_{t}^{\mathbf{B}}x^{d_{M+1}+t-1},\label{FPMM:encodingB} \IEEEeqnarraynumspace
\end{IEEEeqnarray}
where $\mathbf{Z}_{t}^{\mathbf{A}},t\in[K+T_{\mathbf{A}}-1]$ and $\mathbf{Z}_{t}^{\mathbf{B}},t\in[K+T_{\mathbf{B}}-1]$ are 
matrices over $\mathbb{F}$ with the same dimensions as $\mathbf{A}_{\ell,k}^{(\theta_{\mathbf{A}})}$ and $\mathbf{B}_{k,m}^{(\theta_{\mathbf{B}})}$ respectively, whose forms will be specified later.
The product of the two polynomials $f'(x)$ and $h'(x)$ is given by
\begin{IEEEeqnarray}{c}\label{FPMM:productpoly}
g'(x)=\sum\limits_{r=0}^{\delta'}\mathbf{G}_r' x^r=f'(x)\cdot h'(x)
\end{IEEEeqnarray}
with a degree 
\begin{IEEEeqnarray}{l}\label{FPMM:defdelta}
\delta'\!=\!\max\{b_{\ell}\!+\!K\!-\!1,b_{L+1}\!+\!K\!+\!T_{\mathbf{A}}\!-\!2:\ell\!\in\![L]\}\notag\\
\quad\quad +\!\max\{d_m\!+\!K\!-\!1,d_{M+1}\!+\!K\!+\!T_{\mathbf{B}}\!-\!2:m\!\in\![M]\}. \IEEEeqnarraynumspace
\end{IEEEeqnarray}
The degree parameters $\{b_{\ell},d_m:\ell\in[L+1],m\in[M+1]\}$ are said to be \emph{$\delta'$-achievable} for the FPMM problem if the product polynomial $g'(x)$ contains all the sub-matrices in $\mathbf{C}^{(\theta_{\mathbf{A}},\theta_{\mathbf{B}})}=\mathbf{A}^{(\theta_{\mathbf{A}})}\mathbf{B}^{(\theta_{\mathbf{B}})}$ \eqref{FPMM:desiredPro} as coefficients, i.e., $\mathbf{C}_{\ell,m}^{(\theta_{\mathbf{A}},\theta_{\mathbf{B}})}\in\{\mathbf{G}_r':r\in[0:\delta]\}$ for all $\ell\in[L],m\in[M]$.

Next, we use the constructed encoding polynomials with $\delta'$-achievable degree parameters to construct an FPMM strategy with recovery threshold $P=\delta'+1$. To keep the index $\theta_{\mathbf{A}}$ private from any $T_{\mathbf{A}}$ servers and the index $\theta_{\mathbf{B}}$ private from any $T_{\mathbf{B}}$ servers, let $\{\tilde{z}^{(r)}_{\ell,t}:t\in[T_{\mathbf{A}}],\ell\in[L],r\in[R]\}$ and $\{z^{(v)}_{m,t}:t\in[T_{\mathbf{B}}],m\in[M],v\in{V}\}$ be random noises chosen uniformly i.i.d. from $\mathbb{F}$. Then, for all $\ell\in[L],r\in[R]$ and $m\in[M],v\in[V]$, the master constructs the query polynomials $\tilde{q}_{\ell}^{(r)}(x)$ and $q_{m}^{(v)}(x)$  based on the structure of the encoding functions in \eqref{FPMM:encodingA} and \eqref{FPMM:encodingB} as
\begin{IEEEeqnarray}{rCll}
\tilde{q}_{\ell}^{(r)}(x)&=&\sum\limits_{t=1}^{T_{\mathbf{A}}}\tilde{z}^{(r)}_{\ell,t}\cdot x^{b_{L+1}+t-1}+\left\{
\begin{array}{@{}ll}
x^{b_{\ell}}, &\mathrm{if}\,\, r=\theta_{\mathbf{A}}\\
0, & \mathrm{if}\,\, r\neq\theta_{\mathbf{A}}
\end{array}
\right.,  \label{FPMM:query:poly2}\IEEEeqnarraynumspace\\
q_{m}^{(v)}(x)&=&\sum\limits_{t=1}^{T_{\mathbf{B}}}z^{(v)}_{m,t}\cdot x^{d_{M+1}+t-1}+\left\{
\begin{array}{@{}ll}
x^{d_{m}}, &\mathrm{if}\,\, v=\theta_{\mathbf{B}}\\
0, & \mathrm{if}\,\, v\neq\theta_{\mathbf{B}}
\end{array}
\right.. \label{FPMM:query:poly2234}\IEEEeqnarraynumspace
\end{IEEEeqnarray}

The master shares the evaluations of these query polynomials at point $x=\alpha_i$ with server $i$:
\begin{IEEEeqnarray}{rCl}
\mathcal{Q}_{i}^{(\theta_{\mathbf{A}})}&=&\{\tilde{q}_{\ell}^{(r)}(\alpha_i):\ell\in[L],r\in[R]\},\label{FPMM:queryA}\\
\mathcal{Q}_{i}^{(\theta_{\mathbf{B}})}&=&\{q_{m}^{(v)}(\alpha_i):m\in[M],v\in[V]\}. \label{FPMM:queryB}
\end{IEEEeqnarray}

According to the received query $\mathcal{Q}_{i}^{(\theta_{\mathbf{A}})}$, server $i$ encodes its stored data $\mathcal{E}_{i}^{\mathbf{A}}$ \eqref{PSMM:StorageA} into
\begin{IEEEeqnarray}{rCl}
\widetilde{\mathbf{A}}_i&\!=\!&\sum\limits_{r=1}^{R}\sum\limits_{\ell=1}^{L}\tilde{q}_{\ell}^{(r)}(\alpha_i)\cdot \tilde{e}_{\ell}^{(r)}(\alpha_i)\label{FPMM:answerA}\\
&\!\overset{(a)}{=}\!&\sum\limits_{\ell=1}^{L}\left(\sum\limits_{k=1}^{K}\mathbf{A}_{\ell,k}^{(\theta_{\mathbf{A}})}\alpha_i^{k-1}\right)\alpha_i^{b_{\ell}}\notag\\
&&+\!\sum\limits_{r=1}^{R}\sum\limits_{\ell=1}^{L}\!
\left(\sum\limits_{k=1}^{K}\mathbf{A}_{\ell,k}^{(r)}\alpha_i^{k-1}\!\right)\!\!\!\left(\sum\limits_{t=1}^{T_{\mathbf{A}}}\tilde{z}^{(r)}_{\ell,t}\cdot \alpha_i^{b_{L+1}+t-1}\!\right)\label{FSMM:expanding} \IEEEeqnarraynumspace\\
&\!=\!&\sum\limits_{\ell=1}^{L}\left(\sum\limits_{k=1}^{K}\mathbf{A}_{\ell,k}^{(\theta_{\mathbf{A}})}\alpha_i^{k-1}\right)\alpha_i^{b_{\ell}}+\sum\limits_{t=1}^{K+T_{\mathbf{A}}-1}\mathbf{Z}_{t}^{\mathbf{A}}\alpha_i^{b_{L+1}+t-1} \notag\\
&\!\overset{(b)}{=}\!&f'(\alpha_i),\label{FPMM:encodedingA}
\end{IEEEeqnarray}
where $(a)$ is due to \eqref{FPMM:stored_codeA} and \eqref{FPMM:query:poly2}, $(b)$ follows by \eqref{FPMM:encodingA}, and $\mathbf{Z}_{t}^{\mathbf{A}},t\in[K+T_{\mathbf{A}}-1]$ represent various linear combinations of interference sub-matrices $\mathbf{A}_{\ell,k}^{(r)},\ell\in[L],k\in[K],r\in[R]$ that can be found explicitly by expanding the second term in \eqref{FSMM:expanding}, but whose exact forms are unimportant.

Similarly, upon receiving $\mathcal{Q}_{i}^{(\theta_{\mathbf{B}})}$,  by \eqref{FPMM:stored_codeA}, \eqref{FPMM:query:poly2234} and \eqref{FPMM:encodingB}, the stored data $\mathcal{E}_{i}^{\mathbf{B}}$ \eqref{PSMM:StorageB} is encoded into
\begin{IEEEeqnarray}{rCl}
\widetilde{\mathbf{B}}_i&\!=\!&\sum\limits_{v=1}^{V}\sum\limits_{m=1}^{M}q_{m}^{(v)}(\alpha_i)\cdot e_m^{(v)}(\alpha_i)\label{FPMM:answerB}\\
&\!=\!&\sum\limits_{m=1}^{M}\left(\sum\limits_{k=1}^{K}\mathbf{B}_{k,m}^{(\theta_{\mathbf{B}})}\alpha_i^{K-k}\right)\alpha_i^{d_{m}}\notag\\
&\!\!&+\!\sum\limits_{v=1}^{V}\sum\limits_{m=1}^{M}\!\!\left(\sum\limits_{k=1}^{K}\!\mathbf{B}_{k,m}^{(v)}\alpha_i^{K-k}\!\right)\!\!\!\left(\sum\limits_{t=1}^{T_{\mathbf{B}}}z^{(v)}_{m,t}\!\cdot\! \alpha_i^{d_{M+1}+t-1}\!\right)\label{FSMM:expanding2}\IEEEeqnarraynumspace\\
&\!=\!&\sum\limits_{m=1}^{M}\!\left(\sum\limits_{k=1}^{K}\mathbf{B}_{k,m}^{(\theta_{\mathbf{B}})}\alpha_i^{K-k}\!\right)\!\alpha_i^{d_{m}}\!+\!\sum\limits_{t=1}^{K+T_{\mathbf{B}}-1}\mathbf{Z}_{t}^{\mathbf{B}}\alpha_i^{d_{M+1}+t-1} \notag\\
&\!=\!&h'(\alpha_i),\label{FPMM:encodedingB}
\end{IEEEeqnarray}
where $\mathbf{Z}_{t}^{\mathbf{B}},t\in[K+T_{\mathbf{B}}-1]$ can be obtained explicitly by expanding the second term in \eqref{FSMM:expanding2}, but whose exact forms are irrelevant.
Notably, both $\mathbf{Z}_{t}^{\mathbf{A}},t\in[K+T_{\mathbf{A}}-1]$ and $\mathbf{Z}_{t}^{\mathbf{B}},t\in[K+T_{\mathbf{B}}-1]$ are independent of the servers and thus can be viewed as constant terms.

Each server $i$ computes the product $\mathbf{Y}_i^{(\theta_{\mathbf{A}},\theta_{\mathbf{B}})}=\widetilde{\mathbf{A}}_i\widetilde{\mathbf{B}}_i$ and send it back to the master, which is equivalent to evaluating the product polynomial $g'(x)=f'(x)\cdot h'(x)$ at point $x=\alpha_i$ by \eqref{FPMM:encodedingA} and \eqref{FPMM:encodedingB}.
Thus, the master can interpolate the product $g'(x)$ from the responses of any $P=\deg(g'(x))+1=\delta'+1$ servers by \eqref{FPMM:defdelta}, and then recovers the desired computation $\mathbf{C}^{(\theta_{\mathbf{A}},\theta_{\mathbf{B}})}$ by the $\delta'$-achievability of the selected degree parameters for the FPMM problem.

In the following lemma, we present three explicit choices of degree parameters \cite{Zhu_SDMM,EP_SMC} for the FPMM problem. Similar to Lemma \ref{choose:degree} and \eqref{expanding}, it is easy to prove the achievability of these degree parameters, by expanding the product polynomial $g'(x)=f'(x)\cdot h'(x)$ in \eqref{FPMM:productpoly}.
\begin{Lemma}\label{FPMM:acheva:lemma}
The following three explicit choices of degree parameters are achievable for the FPMM problem.
\begin{itemize}
  \item $\delta_1'=(L+1)(KM+K+T_{\mathbf{B}}-1)+T_{\mathbf{A}}-T_{\mathbf{B}}-2$ by setting $b_{\ell}\!=\!(\ell-1)(KM+K+T_{\mathbf{B}}-1),b_{L+1}\!=\!(L-1)(KM+K+T_{\mathbf{B}}-1)+KM,d_m\!=\!(m\!-\!1)K,d_{M\!+\!1}\!\!=\!\!K\!M$ for all $\ell\!\in\![L],m\!\in\![M]$;
  \item $\delta_2'=(M+1)(LK+K+T_{\mathbf{A}}-1)+T_{\mathbf{B}}-T_{\mathbf{A}}-2$ by setting $b_{\ell}=(\ell-1)K,b_{L+1}=LK,d_m=(m-1)(LK+K+T_{\mathbf{A}}-1),d_{M+1}=(M-1)(LK+K+T_{\mathbf{A}}-1)+LK$ for all $\ell\in[L],m\in[M]$;
  \item and $\delta_3'=2LKM+2K+T_{\mathbf{A}}+T_{\mathbf{B}}-4$ by setting $b_{\ell}=(\ell-1)MK,b_{L+1}=LKM,d_m=(m-1)K,d_{M+1}=LKM$ for all $\ell\in[L],m\in[M]$.
\end{itemize}
\end{Lemma}

As a result, our FPMM strategy achieves a recovery threshold of $P=\min\{\delta_1'+1,\delta_2'+1,\delta_3'+1\}$.

\subsection{Privacy and Complexity Analyses}\label{proof:strategy:2}
In this subsection, we prove the privacy of the FPMM strategy and analyze its system performance. 

\begin{Theorem}
The proposed FPMM strategy provides information-theoretical privacy guarantees for the index $\theta_{\mathbf{A}}$ (resp. $\theta_{\mathbf{B}}$) even if any group of up to $T_{\mathbf{A}}$ (resp. $T_{\mathbf{B}}$) servers collude.
\end{Theorem}
\begin{IEEEproof}
We first prove the privacy of the index $\theta_{\mathbf{A}}$ for any $T_{\mathbf{A}}$ colluding servers.
For any subset $\mathcal{T}_{\mathbf{A}}\subseteq[N]$ with $|\mathcal{T}_{\mathbf{A}}|=T_{\mathbf{A}}$, it is straightforward to prove
\begin{IEEEeqnarray}{rCl}
&&I(\theta_{\mathbf{A}};\mathcal{Q}_{\mathcal{T}_{\mathbf{A}}}^{(\theta_{\mathbf{A}})},\mathcal{Q}_{\mathcal{T}_{\mathbf{A}}}^{(\theta_{\mathbf{B}})},\mathcal{E}_{\mathcal{T}_{\mathbf{A}}}^{\mathbf{A}},\mathcal{E}_{\mathcal{T}_{\mathbf{A}}}^{\mathbf{B}},\mathbf{Y}_{\mathcal{T}_{\mathbf{A}}}^{(\theta_{\mathbf{A}},\theta_{\mathbf{B}})})\notag\\
&\overset{(a)}{=}&I(\theta_{\mathbf{A}};\mathcal{Q}_{\mathcal{T}_{\mathbf{A}}}^{(\theta_{\mathbf{A}})},\mathcal{Q}_{\mathcal{T}_{\mathbf{A}}}^{(\theta_{\mathbf{B}})})
\overset{(b)}{=}I(\theta_{\mathbf{A}};\mathcal{Q}_{\mathcal{T}_{\mathbf{A}}}^{(\theta_{\mathbf{A}})})\notag\\
&\overset{(c)}{=}&I(\theta_{\mathbf{A}};\{\tilde{q}_{\ell}^{(r)}\!(\alpha_i):i\in\mathcal{T}_{\mathbf{A}}\}_{\ell\in[L],r\in[R]})
\overset{(d)}{=}0, \notag
\end{IEEEeqnarray}
where $(a)$ is similar to \eqref{PSMM:proof:privacy}; 
$(b)$ is because $\mathcal{Q}_{\mathcal{T}_{\mathbf{A}}}^{(\theta_{\mathbf{B}})}$ \eqref{FPMM:queryB} is generated independently of $\theta_{\mathbf{A}}$ and $\mathcal{Q}_{\mathcal{T}_{\mathbf{A}}}^{(\theta_{\mathbf{A}})}$ \eqref{FPMM:queryA} by \eqref{FPMM:query:poly2}-\eqref{FPMM:query:poly2234};
$(c)$ follows by \eqref{FPMM:queryA}; $(d)$ is due to Lemma \ref{security proof} and the fact that the query polynomial $\tilde{q}_{\ell}^{(r)}(x)$ is constructed by using $T_{\mathbf{A}}$ random noises to mask the desired index $\theta_{\mathbf{A}}$ by \eqref{FPMM:query:poly2} for all $\ell\in[L],r\in[R]$, and all these noises are i.i.d. uniformly distributed on $\mathbb{F}$. Thus the privacy of index $\theta_{\mathbf{A}}$ follows by \eqref{security:AAA}. 

Similarly, it is also straightforward to prove that the privacy of index $\theta_{\mathbf{B}}$ satisfies the constraint in \eqref{security:BBB}.
\end{IEEEproof}

\begin{Theorem}\label{theorem:FPMM}
For the FPMM problem with $(N,K)$ MDS-coded storage and colluding parameters $T_{\mathbf{A}},T_{\mathbf{B}}$, let $L,M$ be arbitrary partitioning parameters, then the proposed FPMM strategy can achieve
\begin{IEEEeqnarray*}{l}
\text{Recovery Threshold:}\\ 
\;\; P=\min\big\{(L+1)(KM+K+T_{\mathbf{B}}-1)+T_{\mathbf{A}}-T_{\mathbf{B}}-1,\\
\;\;\quad\quad\quad\quad\;\; (M+1)(LK+K+T_{\mathbf{A}}-1)+T_{\mathbf{B}}-T_{\mathbf{A}}-1,\\
\;\;\quad\quad\quad\quad\;\; 2LKM+2K+T_{\mathbf{A}}+T_{\mathbf{B}}-3\big\}, \\
\text{Download Cost:}\;\; D=\frac{\lambda\gamma P}{LM},\\
\text{Server Computation:} \;\; {C}_{s}={O}\left(\frac{R\lambda\omega+V\omega\gamma}{K}+\frac{\lambda\omega\gamma}{LKM}\right), \\
\text{Decoding Complexity:} \;\; {C}_d={{O}}\left(\frac{\lambda\gamma P(\log P)^2\log\log P}{LM}\right).
\end{IEEEeqnarray*}
\end{Theorem}
\begin{IEEEproof}
Section \ref{proof:theorem:FPMM} tells us that the proposed FPMM strategy achieves the recovery threshold.

In terms of download cost, the master downloads a matrix with the dimension of $\frac{\lambda}{L}\times\frac{\gamma}{M}$ from each of responsive servers, and thus the strategy achieves a download cost of $D=\frac{\lambda\gamma P}{LM}$.

The computation at each server includes encoding the matrices $\mathbf{A}^{([R])}$ by taking a linear combination of $RL$ sub-matrices with dimension of $\frac{\lambda\omega}{LK}$ \eqref{FPMM:answerA}, then encoding the matrices $\mathbf{B}^{([V])}$ by taking a linear combination of $VM$ sub-matrices with dimension of $\frac{\omega\gamma}{KM}$ \eqref{FPMM:answerB}, and finally multiplying the two encoded sub-matrices with sizes of $\frac{\lambda}{L}\times\frac{\omega}{K}$ and $\frac{\omega}{K}\times\frac{\gamma}{M}$, which require a complexity of ${O}(\frac{R\lambda\omega}{K}+\frac{V\omega\gamma}{K}+\frac{\lambda\omega\gamma}{LKM})$ if we employ straightforward matrix multiplication algorithms.
Decoding requires interpolating a polynomial of degree $P-1$ from $P$ arbitrary responses for $\frac{\lambda\gamma}{LM}$ times, which achieves the complexity ${{O}}(\frac{\lambda\gamma P(\log P)^2\log\log P}{LM})$ by Lemma \ref{com:poly}.
\end{IEEEproof}

\begin{Remark}\label{MDS:effect}
Here we compare the differences between the two proposed strategies and show the effect of MDS-coded storage on them.
Both the proposed PSMM and FPMM strategies are constructed by first designing encoding polynomials of the desired matrix multiplication, and then exploiting the structures inspired by the encoding polynomials to create secret shares of confidential matrix and/or private indices, such that the desired computation can be completed by interpolating a polynomial from server responses. When MDS-coded storage is considered, the main challenge lies in how to design these encoding polynomials that are compatible with the given MDS-coded storage structure, such that the encoding functions can be exploited to design private queries and the interference from undesired matrices are aligned along low dimensions as much as possible. More specifically,  for the proposed PSMM strategy, when the MDS-coded storage in \eqref{storage:PSMM} is given,  the encoding polynomial of matrix $\mathbf{B}^{(\theta)}$ is designed to match the MDS-coded structure. Furthermore, our private queries can align the interference from the undesired matrices $\mathbf{B}^{([N]\backslash\{\theta\})}$ along the $K+T-1$ dimensions $x^{d_{M+1}},x^{d_{M+1}+1},\ldots,x^{d_{M+1}+K+T-2}$ by \eqref{encoding:alignment}, thus the encoding polynomial of matrix $\mathbf{B}^{(\theta)}$ includes these dimensions for interference alignment. To make this more clear, we rewrite the encoding function \eqref{PSMM:encodingB} of $\mathbf{B}^{(\theta)}$ as follow: 
\begin{IEEEeqnarray}{c}\label{end:stru}
h(x)\!\!=\!\!\sum\limits_{m=1}^{M}\!\!\Big(\underbrace{\sum\limits_{k=1}^{K}\mathbf{B}_{k,m}^{(\theta)}x^{K\!-\!k}}_{\text{MDS-coded Structure}}\Big)x^{d_m}
\!+\!\underbrace{\sum\limits_{t=1}^{K\!+\!T\!-\!1} \!\!\!\mathbf{Z}_{t}^{\mathbf{B}}x^{d_{M\!+\!1}+t-1}}_{\text{Interference Alignment}}. \IEEEeqnarraynumspace
\end{IEEEeqnarray}
Accordingly, the encoding function \eqref{PSMM:encodingA} of the confidential matrix $\mathbf{A}$ also needs to match the MDS-coded storage structure for completing the desired computation $\mathbf{A}\mathbf{B}^{(\theta)}$.
However, the matrix $\mathbf{A}$ is owned by the master and there is no interference from other undesired matrices. Thus we just use $S$-dimensional noises to ensure secrecy of $\mathbf{A}$:
\begin{IEEEeqnarray}{c}\notag
f(x)\!=\!\sum\limits_{\ell=1}^{L}\!\!\Big(\underbrace{\sum\limits_{k=1}^{K}\mathbf{A}_{\ell,k}x^{k-1}}_{\text{MDS-coded Structure}}\Big)x^{b_{\ell}}
+\underbrace{\sum\limits_{t=1}^{S}\mathbf{Z}_{t}^{\mathbf{A}}x^{b_{L+1}+t-1}}_{\text{Mask Noises}}.
\end{IEEEeqnarray}
In the FPMM strategy, the desired matrices $\mathbf{A}^{(\theta_{\mathbf{A}})}$ and $\mathbf{\mathbf{B}}^{(\theta_{\mathbf{B}})}$ are stored in MDS-coded forms, similar to $\mathbf{B}^{(\theta)}$ in the PSMM strategy. Thus the encoding polynomials for the two matrices  are constructed following a similar structure to \eqref{end:stru}, see \eqref{FPMM:encodingA} and \eqref{FPMM:encodingB}. 
In general, the main differences between the two strategies of PSMM and  FPMM lie in designing these encoding polynomials of desired matrix multiplication,  i.e.,
\begin{itemize}
    \item The encoding polynomials of the desired matrices $\mathbf{\mathbf{B}}^{(\theta)},\mathbf{A}^{(\theta_{\mathbf{A}})}, \mathbf{\mathbf{B}}^{(\theta_{\mathbf{B}})}$ are constructed following the similar coding structure, which needs to match the MDS-coded storage structure and aligns the interference from undesired matrices.
    \item The encoding polynomial of the matrix $\mathbf{A}$ owned by the master is constructed without any interference from undesired matrices, which just needs to match the MDS-coded structure and ensures the secrecy of $\mathbf{A}$.
\end{itemize}
\end{Remark}

\begin{Remark}\label{com:reviso}
Table \ref{tab:com} compares the performance between the proposed PSMM and FPMM strategies. In general, the two strategies achieve almost the same communication and computation costs. Particularly,  the recovery thresholds of the two strategies are equal by setting $S=K+T_{\mathbf{A}}-1$ and $T=T_{\mathbf{B}}$, which means that the performance of the two strategies are almost identical by replacing the confidential matrix $\mathbf{A}$ (that is secure against $S$ servers and is owned by the master) with the private matrix $\mathbf{A}^{(\theta_{\mathbf{A}})}$ (that is private against $T_{\mathbf{A}}$ servers and is stored at the distributed system in an $(N,K)$ MDS code).

\begin{table}[htbp]
\centering
\caption{Performance comparison between the proposed PSMM and FPMM strategies.} \label{tab:com}
\resizebox{87mm}{16mm}{
  \begin{tabular}{|@{\,}c@{\,}|@{}c@{}|@{}c@{}|}
  \hline
  & Our PSMM Strategy  & Our FPMM Strategy  \\ \hline
Recovery Threshold & $P$ & $P'$  \\ \hline
  Upload Cost & $\frac{\lambda\omega N}{LK}$ &  $\backslash$  \\ \hline
  Download Cost & $\frac{\lambda\gamma P}{LM}$ &  $\frac{\lambda\gamma P'}{LM}$  \\ \hline
  Encoding Comp. & $O(\frac{\lambda\omega N(\log N)^2\log\log N}{LK})$ & $\backslash$ \\ \hline
  Server Comp. & ${O}(\frac{V\omega\gamma}{K}+\frac{\lambda\omega\gamma}{LKM})$ & ${O}(\frac{R\lambda\omega}{K}+\frac{V\omega\gamma}{K}+\frac{\lambda\omega\gamma}{LKM})$ \\ \hline
  Decoding Comp. & ${{O}}(\frac{\lambda\gamma P(\log P)^2\log\log P}{LM})$ & ${{O}}(\frac{\lambda\gamma P'(\log P')^2\log\log P'}{LM})$  \\ \hline
  \end{tabular}
  }
\begin{tablenotes}
       \footnotesize
       \item[] Here $P\!=\!\min\big\{2LKM\!+\!K\!+\!S\!+\!T\!-\!2,(M\!+\!1)(LK\!+\!S)\!+\!K\!+\!T\!-\!S\!-\!2,(L\!+\!1)(K\!M\!+\!K\!+\!T\!-\!1)\!+\!S\!-\!K\!-\!T\big\}$ and $P'\!=\!\min\big\{2LKM\!+\!2K\!+\!T_{\mathbf{A}}\!+\!T_{\mathbf{B}}\!-\!3,(M\!+\!1)(LK\!+\!K\!+\!T_{\mathbf{A}}\!-\!1)\!+\!T_{\mathbf{B}}\!-\!T_{\mathbf{A}}\!-\!1,(L\!+\!1)(K\!M\!+\!K\!+\!T_{\mathbf{B}}\!-\!1)+T_{\mathbf{A}}\!-\!T_{\mathbf{B}}\!-\!1\big\}$.
\end{tablenotes}
\end{table}
\end{Remark}

\section{Numerical Evaluations and Comparisons with Related Works}\label{Comparison}



\subsection{Comparison for PSMM with MDS Coded Storage}\label{subsection:MDScom}
The problem of PSMM with MDS coded storage studied in \cite{PSMM:3} is most closely related to the PSMM problem considered in this paper, which includes the PSMM problem in \cite{PSMM:3} as a special case with $S=T=1$. A computation strategy from $(N,K=pn)$ MDS-coded storage for some fixed parameters $p,n$ was constructed in~\cite{PSMM:3}.

In essence, both the previous strategy in \cite{PSMM:3} and our PSMM strategy employ polynomial codes \cite{Polynomial_code,EP_code} to encode the confidential matrix $\mathbf{A}$ in conjunction with using the idea of Shamir's secret sharing \cite{Shamir} to keep $\mathbf{A}$ secure. 
The key difference lies in how the private queries are designed.
In~\cite{PSMM:3}, to privately retrieve the desired matrix multiplication $\mathbf{A}\mathbf{B}^{(\theta)}$ from the distributed system storing $V$ matrices, the queries are designed such that 1) the query sent to each server consists of $V$ matrices of dimension $\frac{\lambda\omega}{Lp}$ each, and all the query matrices are uniformly and independently distributed on $\mathbb{F}$ in terms of any individual server, which ensures the privacy of the desired index $\theta$; 2) the query matrices corresponding to the desired index $\theta$ are the evaluations of encoding function of $\mathbf{A}$ for completing the desired computation, whereas the remaining query matrices are identical across all the servers, which align interference from undesired computation. 
Moreover, in the prior strategy \cite{PSMM:3}, upon receiving  $V$ query matrices from the master, each server computes the pairwise multiplications of the $V$ received matrices and the $V$ stored encoding matrices. Evidently, both the upload cost and server computation cost strictly depend on the number of matrices $V$. For large $V$ in many big data applications, the strategy in \cite{PSMM:3} is usually invalid in terms of communication and computation costs.

Since the interference matrices in the queries \cite{PSMM:3} are identical across all the servers, it is difficult to generalize the query design to the colluding case, while protecting the privacy of the index $\theta$, guaranteeing the correctness of the computation result $\mathbf{A}\mathbf{B}^{(\theta)}$, and ensuring the efficiency of the communication and computation.
In our PSMM strategy, we carefully design encoding functions of the desired matrix multiplication $\mathbf{A}\mathbf{B}^{(\theta)}$ and then exploit these encoding functions to create secret shares of the matrix $\mathbf{A}$ and the private index $\theta$ under the privacy and secrecy constraints (see \eqref{encoding:A} and \eqref{PSMM:queryB}), which are perfectly compatible with the matrix multiplication task and MDS-coded storage structure, such that the computation results from the servers can be viewed as evaluations of a polynomial at distinct points, from which the intended result can be obtained through polynomial interpolation.
Consequently, our strategy sends an encoding matrix of $\mathbf{A}$ with dimension $\frac{\lambda\omega}{Lp}$ and a query of size $V$ to each server by \eqref{encoding:A} and \eqref{PSMM:queryB}. Each server computes a linear combination of its stored encoding matrices using the received query as coefficients \eqref{encoding:B}, and then multiplies the received encoding matrix of $\mathbf{A}$ with the result of the linear combination \eqref{PSMM:responses}, i.e., our strategy just  uploads a matrix and computes a pair of matrix multiplication at each server. Thus our strategy significantly decreases the communication and computation costs.

More specifically, for the special case of $S=T=1$, the complexity performance of the PSMM strategy in \cite{PSMM:3} and our strategy are presented in Table \ref{tab:PSMM}.
Compared with \cite{PSMM:3}, our strategy reduces the upload cost by a factor of $Vn$, the encoding complexity by a factor of $O(n)$, and the server computation complexity by a factor of $O(V\!M)$ if $\frac{V\omega\gamma}{pn}\!<\!\frac{\lambda\omega\gamma}{LMpn}$ or $O(\frac{\lambda}{L})$ otherwise, but increasing the recovery threshold by a factor of $O(M)$, and the download cost and the decoding complexity by a factor of $O(n)$. Moreover, both the two strategies achieve the same storage cost.
%

\begin{table}[htbp]
\centering
\small
\caption{Performance comparison between the strategy \cite{PSMM:3} and our strategy for the special case of $S=T=1$.} \label{tab:PSMM}
\resizebox{87mm}{21mm}{
  \begin{tabular}{|@{}c@{}|@{}c@{}|@{}c@{}|}
  \hline
  &  Previous PSMM Strat. \cite{PSMM:3} & Our PSMM Strat.  \\ \hline
\multirow{2}{*}{Recovery} & \multirow{3}{*}{$P'=Lpn+pn$} &  $P\!=\!\min\{2LMpn\!+\!pn,$  \\
& & $(\!M\!+\!1\!)\!(\!Lpn\!+\!1\!)\!+\!pn\!-\!2,$ \\
Threshold & & $(\!L\!+\!1)\!(Mpn\!+\!pn)\!-\!pn\!\}$   \\ \hline
  Upload Cost & $\frac{\lambda\omega VN}{Lp}$ &  $\frac{\lambda\omega N}{Lpn}$  \\ \hline
  Download Cost & $\frac{\lambda\gamma P'}{Ln}$ &  $\frac{\lambda\gamma P}{LM}$  \\ \hline
  Encoding Comp. & ${{O}}(\frac{\lambda\omega N(\log N)^2\log\log N}{Lp})$ & $O(\frac{\lambda\omega N(\log N)^2\log\log N}{Lpn})$ \\ \hline
  Server Comp. & ${O}(\frac{V\lambda\omega\gamma}{Lpn})$ & ${O}(\frac{V\omega\gamma}{pn}+\frac{\lambda\omega\gamma}{LMpn})$ \\ \hline
  Decoding Comp. & ${{O}}(\frac{\lambda\gamma P'(\log P')^2\log\log P'}{Ln})$ & $O(\frac{\lambda\gamma P(\log P)^2\log\log P}{LM})$  \\ \hline
  Storage Cost  & $\frac{V\omega\gamma}{pn}$ & $\frac{V\omega\gamma}{pn}$  \\ \hline
  \end{tabular}
  }
   \begin{tablenotes}
       \footnotesize
       \item[] Here $p,n$ are some fixed parameters satisfying $K=pn$.
\end{tablenotes}
\end{table}



\begin{figure}[htbp]
	\centering
		\subfigure[$N=50$]
		{ \label{N50}
		\includegraphics[width=0.225\textwidth]{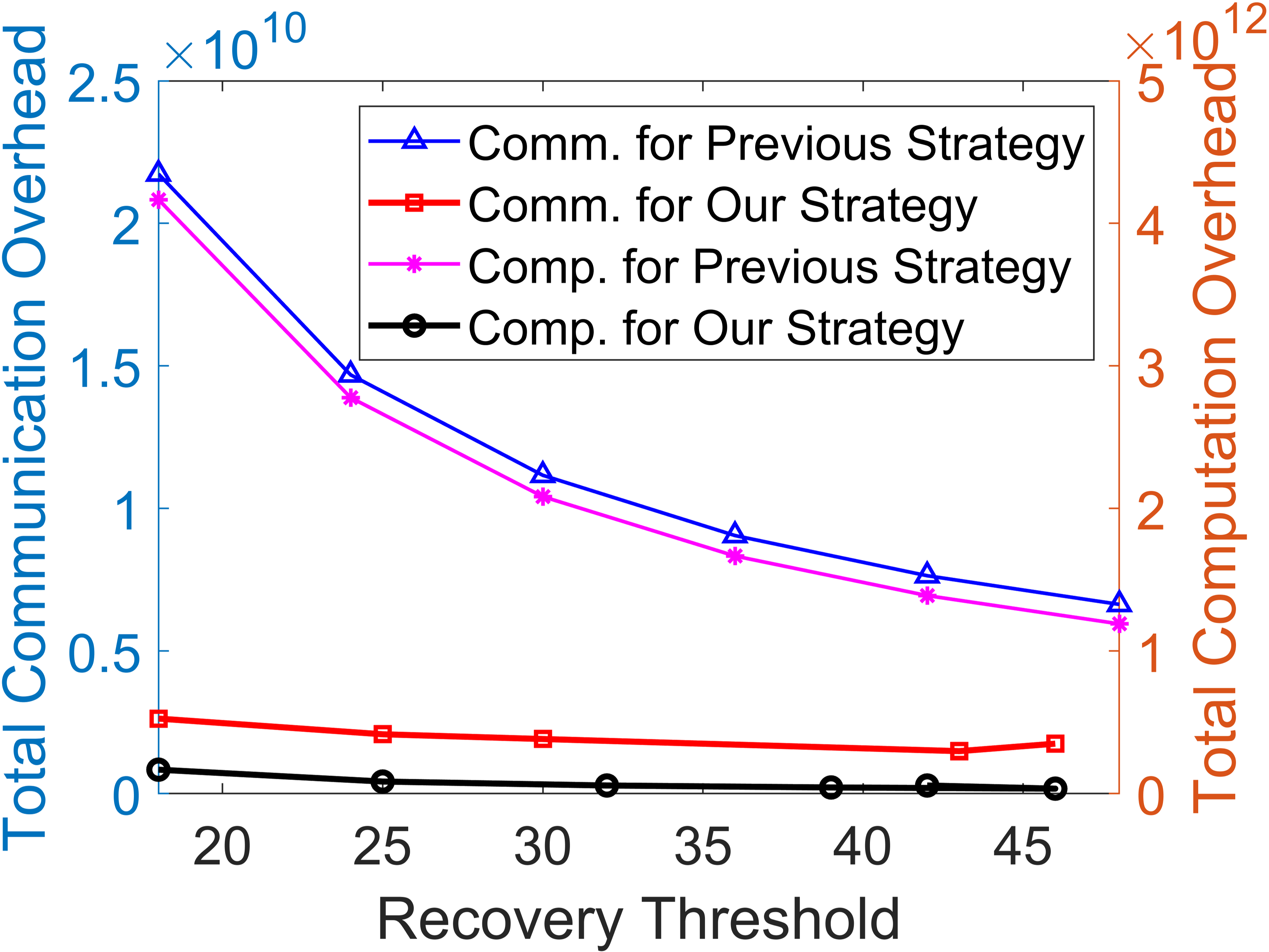}
		} \vspace{-1mm}
		\subfigure[$N=200$]
		{ \label{N200}
		\includegraphics[width=0.225\textwidth]{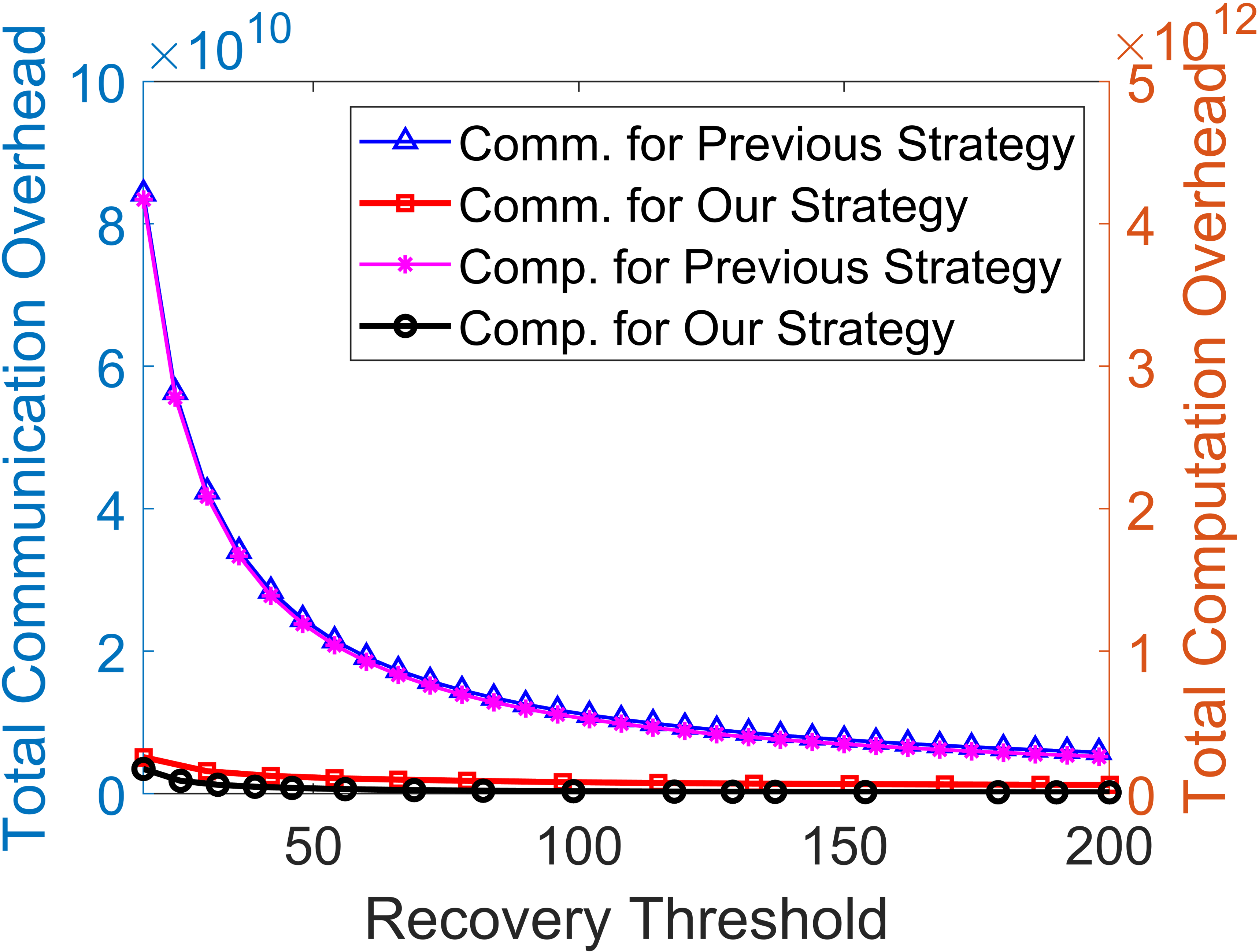}
		}
	\caption{Lower convex hulls of achievable (recovery threshold, total communication overhead) pairs and (recovery threshold, total computation overhead) pairs for the  previous PSMM strategy \cite{PSMM:3} and our PSMM strategy, given $K=6,V=50$ and (a) $N=50$, (b) $N=200$.}
	\label{figure:N}
\end{figure}

\begin{figure}[htbp]
	\centering
		\subfigure[$K=12$]
		{ \label{K12}
		\includegraphics[width=0.225\textwidth]{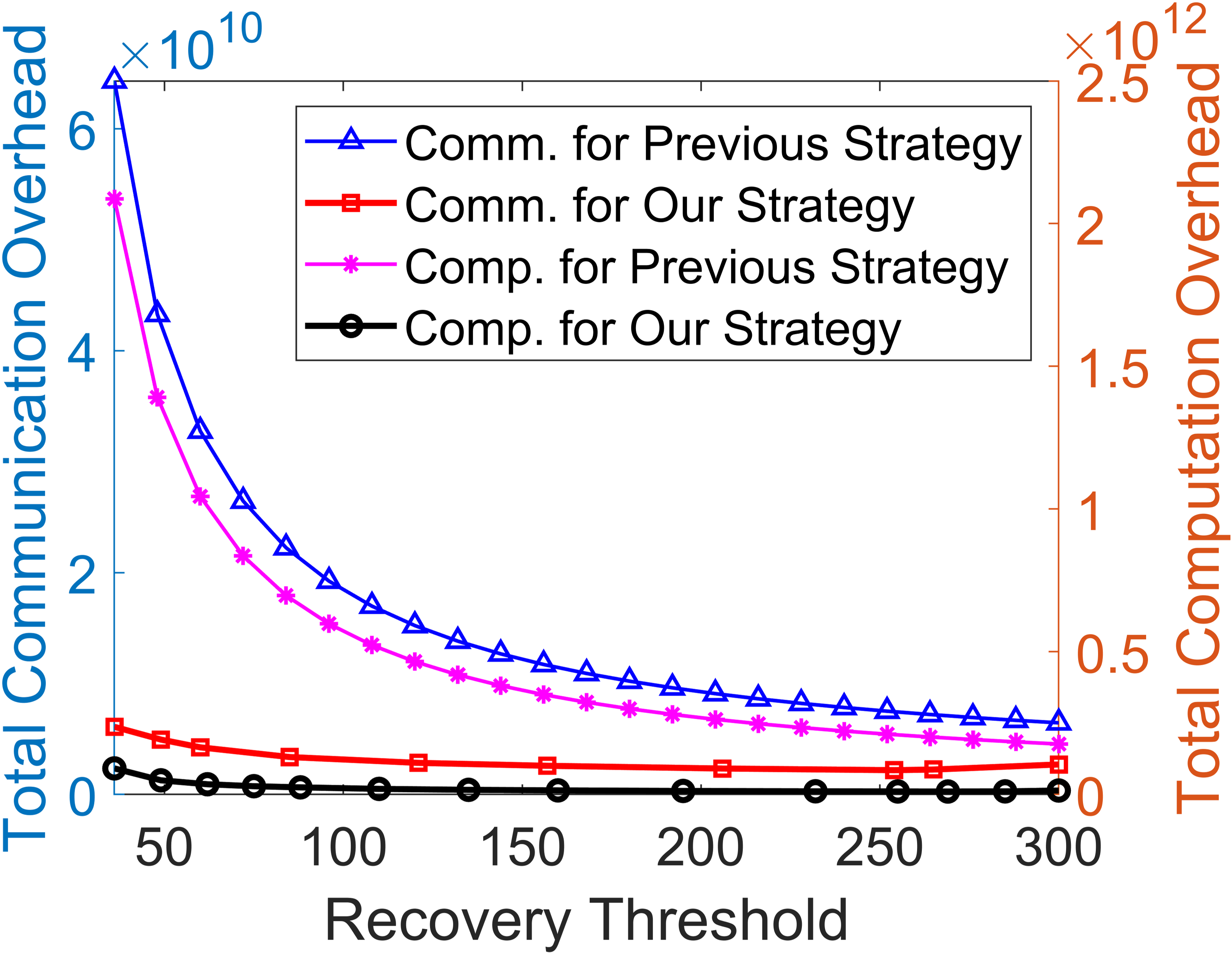}
		}
		\subfigure[$K=42$]
		{ \label{K42}
		\includegraphics[width=0.225\textwidth]{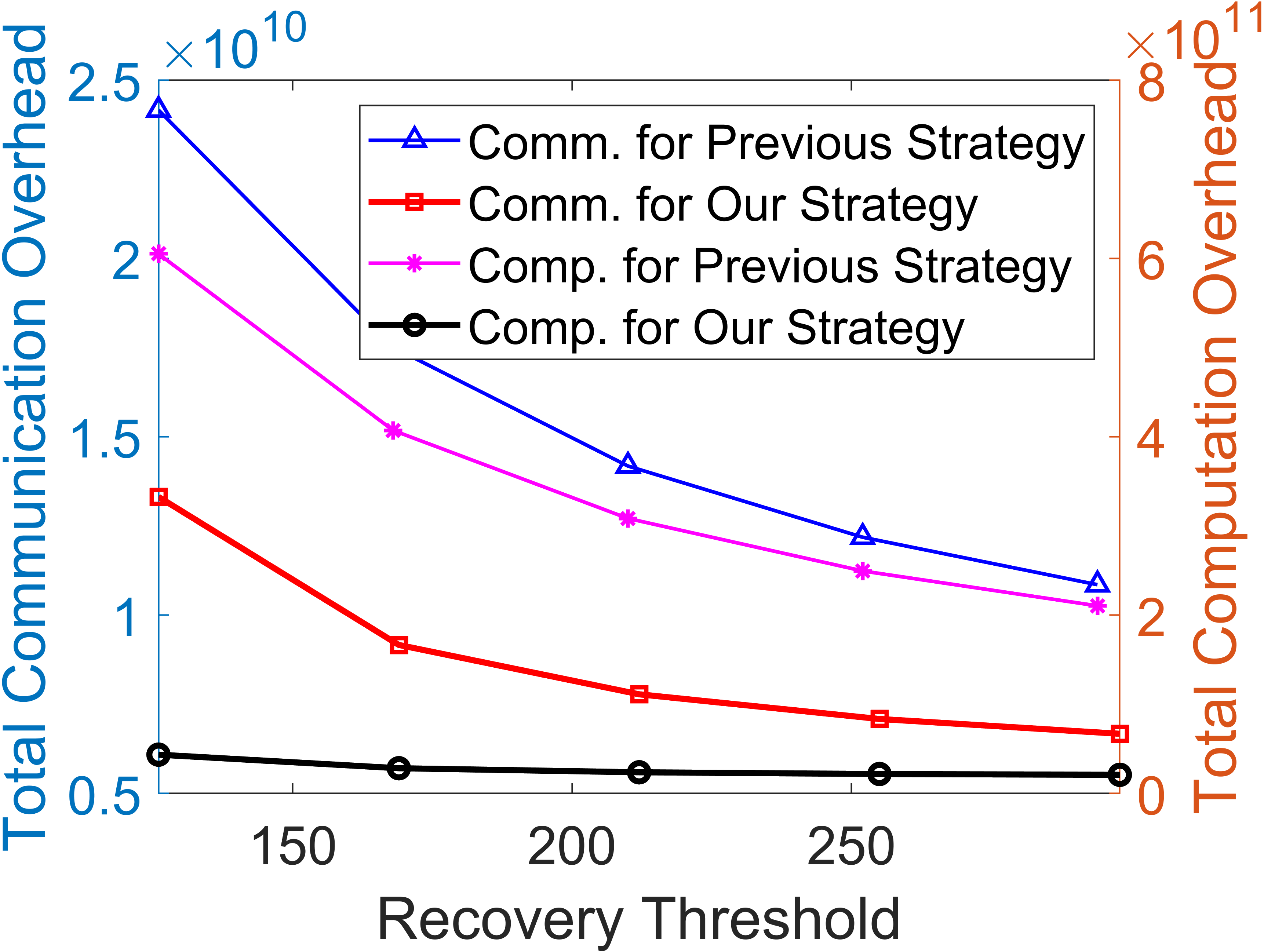}
		}
	\caption{Lower convex hulls of achievable (recovery threshold, total communication overhead) pairs and (recovery threshold, total computation overhead) pairs for the previous PSMM strategy \cite{PSMM:3} and our PSMM strategy, given $V=50,N=300$ and (a) $K=12$, (b) $K=42$.}
	\label{figure:K}
\end{figure}

\begin{figure}[htbp]
	\centering
		\subfigure[$V=50$]
		{ \label{V50}
		\includegraphics[width=0.225\textwidth]{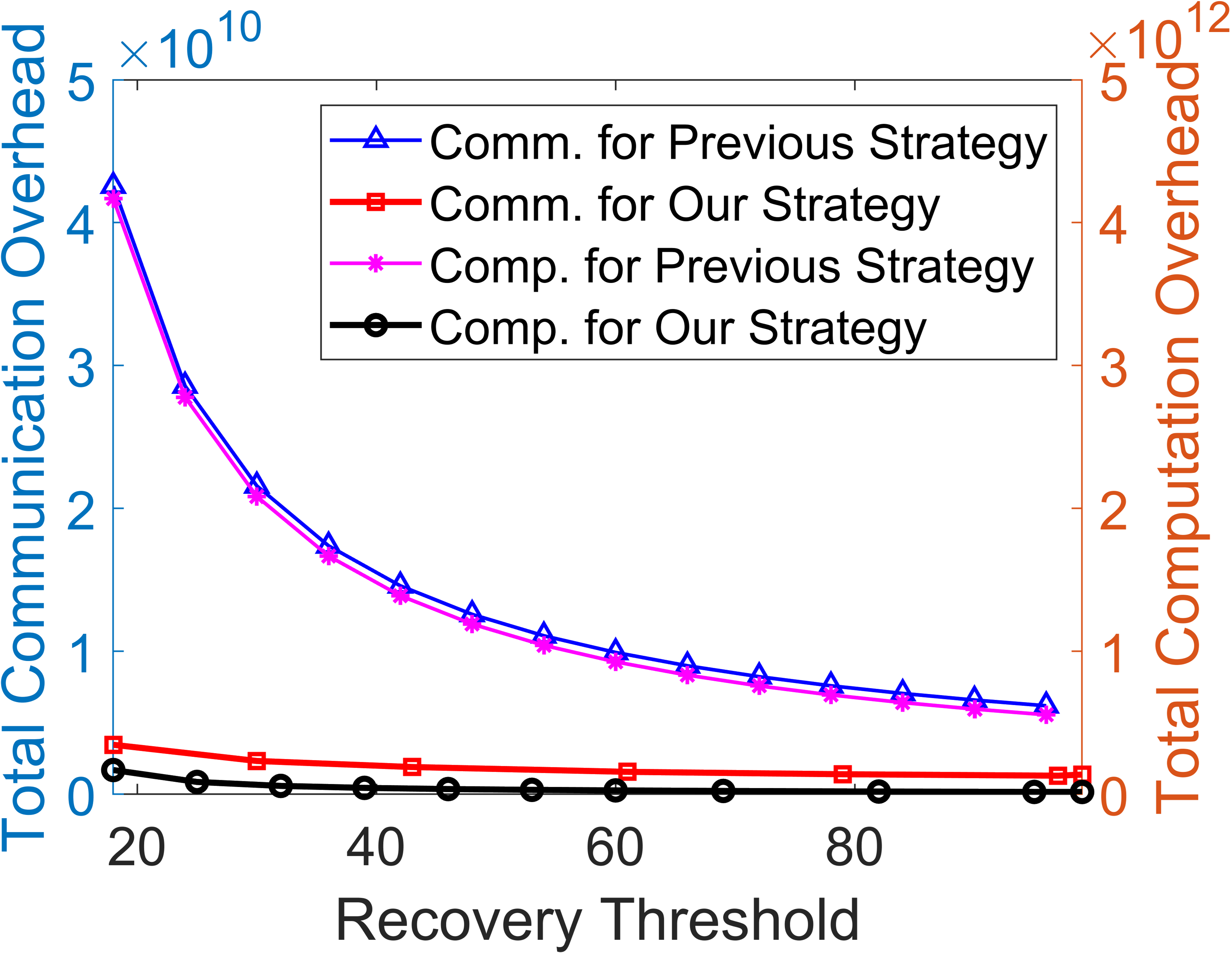}
		}
		\subfigure[$V=200$]
		{ \label{V200}
		\includegraphics[width=0.225\textwidth]{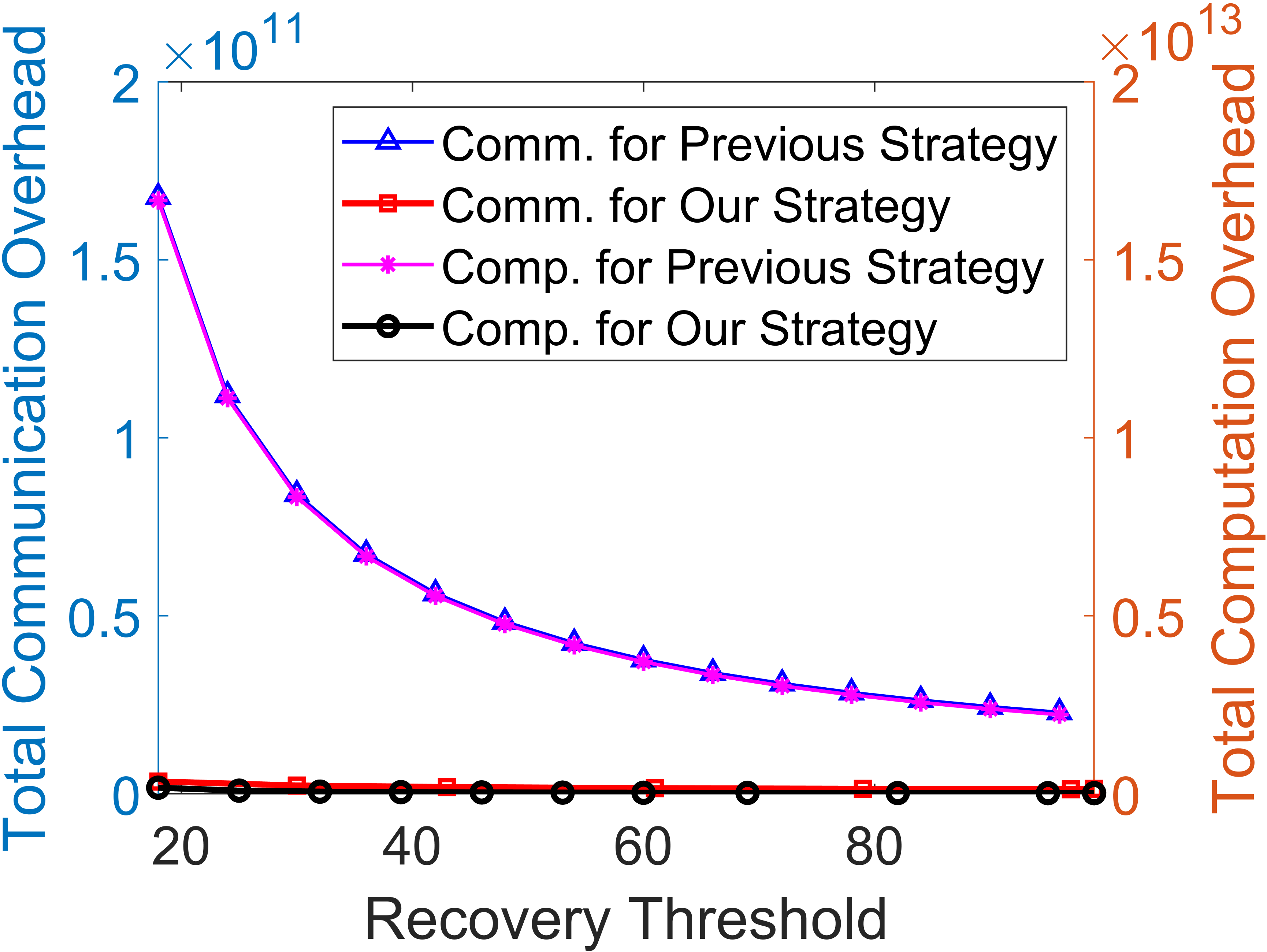}
		}
	\caption{Lower convex hulls of achievable (recovery threshold, total communication overhead) pairs and (recovery threshold, total computation overhead) pairs for the previous PSMM strategy \cite{PSMM:3} and our PSMM strategy, given $K=6,N=100$ and (a) $V=50$, (b) $V=200$.}
	\label{figure:V}
\end{figure}

It is worth noting that, given any fixed MDS-coded parameter $K$, the parameters $L,p,n$ such that $K=pn$ are free in the PSMM strategy in \cite{PSMM:3}, i.e., the previous PSMM strategy  achieves a tradeoff among recovery threshold,  communication and computation overheads by freely varying $L,p,n$ with $K=pn$. Similarly, our  strategy also establishes such a tradeoff via freely adjusting two parameters of $L,M$.
To more clearly compare the complexity performance of our PSMM strategy with the one in \cite{PSMM:3}, we define \emph{total communication overhead} as the sum of upload and download cost, and \emph{total computation overhead} as the sum of encoding, computation and decoding complexities. 
For distinct system parameters $N,K,V$,
Fig. \ref{figure:N}, \ref{figure:K} and \ref{figure:V} show lower convex hulls of achievable (recovery threshold, total communication overhead) pairs and (recovery threshold, total computation overhead) pairs, 
optimized over $L$ and $p,n$ for the previous strategy \cite{PSMM:3} and over $L,M$ for our strategy.  Notably, the parameters $L,M,p,n$ are optimized subject to $K=pn$ and the constraint that the recovery threshold of the corresponding strategy is not more than the number of servers $N$. Moreover, we omit the order notation $O(\cdot)$ to simplify the comparison of the order analysis and set the dimensions of data matrices to be $\lambda=\omega=\gamma=10000$. More specifically, Fig. \ref{N50} and \ref{N200} are obtained by setting $N=50$ and $N=200$, respectively, given $K=6$ and $V=50$. 
Fig. \ref{K12} and \ref{K42} are obtained by setting the MDS-coded  parameter $K=12$ and $K=42$, respectively, given $V=50$ and $N=300$.
Fig. \ref{V50} and \ref{V200} are obtained by setting the number of matrices $V=50$ and $V=200$, respectively, given $K=6$ and $N=100$.

We observe from Fig. \ref{figure:N}, \ref{figure:K} and \ref{figure:V} that
\begin{itemize}
\item For all evaluated combinations of system parameters $N,K,V$, given any recovery threshold, our PSMM strategy exhibits significant advantage over \cite{PSMM:3} in terms of total communication overhead and total computation overhead, particularly for low recovery threshold. Conversely, given the same total communication overhead and total computation overhead required to accomplish the PSMM, our strategy achieves a lower recovery threshold than that in \cite{PSMM:3}. The performance gain of our PSMM strategy over the previous strategy \cite{PSMM:3} comes from the fact that the communication and computation costs in our strategy are almost independent of the system parameter $V$. However, the upload cost and the server computation cost in the previous strategy \cite{PSMM:3} linearly depend on $V$. 
\item As the recovery threshold increases, the total communication and the total computation overheads of the PSMM strategy in \cite{PSMM:3} decrease drastically, while those of our strategy vary slowly (almost remain unchanged). This shows that our strategy has much better robustness against system heterogeneity, and it will have a constantly low execution time, regardless of the presence of slow servers.  
\end{itemize}

\subsection{Comparison with Baseline FPMM Strategy}
For the FPMM problem, one basic strategy is that each server first computes the product of each pair of  encoding sub-matrices and stores these results, and then the master  uses the idea of PIR \cite{MDS-X-security} to privately retrieve the desired matrix multiplication from the results stored at servers. As a comparison with our strategy, we outline the baseline strategy.  

Each server $i$  computes and stores the following results, which can be precomputed during off-peak hours.  
\begin{IEEEeqnarray}{c}
\mathcal{E}_{i}\!=\!\left\{\tilde{e}^{(r)}(\alpha_i)\cdot e^{(v)}(\alpha_i):r\!\in\![R],v\!\in\![V]\right\}, \quad\forall\,i\in[N],\notag
\end{IEEEeqnarray}
where $\tilde{e}^{(r)}(\alpha_i)$ and $e^{(v)}(\alpha_i)$ defined in \eqref{FPMM:encstore}-\eqref{PSMM:StorageB22} are the encoding sub-matrices of $\mathbf{A}^{(r)}$ and $\mathbf{B}^{(v)}$ stored at server $i$ for all $r\in[R]$ and $v\in[V]$, respectively.

To privately retrieve the desired computation $\mathbf{A}^{(\theta_{\mathbf{A}})}\mathbf{B}^{(\theta_{\mathbf{B}})}$, following the idea of PIR in \cite{MDS-X-security}, the query sent to server $i$ is designed as $\mathcal{Q}_{i}=\{q^{(r,v)}(\alpha_i):r\in[R],v\in[V]\}$, where
\begin{IEEEeqnarray}{c}\label{query:base}
q^{(r,v)}(x)\!=\!\sum\limits_{t=1}^{T'}z^{(r,v)}_{t}x^{t-1}\!+\!\left\{
\begin{array}{@{}l@{\;\;}l@{}}
\frac{1}{x^K}, &\mathrm{if}\,\, (r,v)\!=\!(\theta_{\mathbf{A}},\theta_{\mathbf{B}})\\
0, & \mathrm{if}\,\, (r,v)\!\neq\!(\theta_{\mathbf{A}},\theta_{\mathbf{B}})
\end{array}
\right.. \IEEEeqnarraynumspace
\end{IEEEeqnarray}
Here $T'\triangleq\max\{T_{\mathbf{A}},T_{\mathbf{B}}\}$ is the privacy parameter of the designed queries, and $z_{1}^{(r,v)},\ldots,z_{T'}^{(r,v)}$ are the random noises that are used to mask the indices $\theta_{\mathbf{A}},\theta_{\mathbf{B}}$ of the desired computation.

The server $i$ computes and responds
\begin{IEEEeqnarray}{c}\notag
\mathbf{Y}_i^{(\theta_{\mathbf{A}},\theta_{\mathbf{B}})}=\sum\limits_{r\in[R],v\in[V]}q^{(r,v)}(\alpha_i)\cdot\tilde{e}^{(r)}(\alpha_i)\cdot e^{(v)}(\alpha_i),
\end{IEEEeqnarray}
which can be viewed as the evaluation of the following polynomial $g(x)$ at point $x=\alpha_i$ by \eqref{FPMM:encstore} and \eqref{query:base}.
\begin{IEEEeqnarray}{rCl}
g(x)
&=&\sum\limits_{r\in[R],v\in[V]}q^{(r,v)}(x)\cdot\tilde{e}^{(r)}(x)\cdot e^{(v)}(x) \label{expand:base}\\
&=&\underbrace{\sum\limits_{i=0}^{K-2}\bigg(\sum\limits_{j=0}^{i}\mathbf{A}_{j+1}^{(\theta_{\mathbf{A}})}\mathbf{B}_{K+j-i}^{(\theta_{\mathbf{B}})}\bigg)\frac{1}{x^{K-i}}}_{\text{Interference Terms}}\notag\\
&&+\underbrace{\bigg(\sum\limits_{j=0}^{K-1}\mathbf{A}_{j+1}^{(\theta_{\mathbf{A}})}\mathbf{B}_{j+1}^{(\theta_{\mathbf{B}})}\bigg)\frac{1}{x}}_{\text{Desired Term}}+\underbrace{\sum\limits_{t=0}^{2K+T'-3}\mathbf{Z}_{t}x^{t}}_{\text{Interference Terms}}, \notag
\end{IEEEeqnarray}
where $\mathbf{Z}_{t},t\in[2K+T'-3]$ represent various linear combinations of interference sub-matrices and can be obtained explicitly by expanding the term in \eqref{expand:base}.

By the invertibility of the Vandermonde matrix,  the master can recover the polynomial $g(x)$ from any $P=3K+T'-2$ responses, and then obtain the desired computation due to $\mathbf{A}^{(\theta_{\mathbf{A}})}\mathbf{B}^{(\theta_{\mathbf{B}})}=\sum_{j=0}^{K-1}\mathbf{A}_{j+1}^{(\theta_{\mathbf{A}})}\mathbf{B}_{K+j-1}^{(\theta_{\mathbf{B}})}$ by \eqref{FPMM:partition}. 

In Table \ref{baseline}, we summarize the performance of the baseline strategy. Compared with our FPMM strategy, the baseline strategy achieves a slightly lower recovery threshold, download cost and 
decoding complexity, and reduces the server computation complexity  by a factor of $O(\frac{\omega}{RV})$, but the storage cost is increased by a factor of $O(K\cdot\min\{R,V\})$, where $V$ and $R$ are the number of matrices and are typically large in the current era of big data.

\begin{table}[htbp]
\centering
\small
\caption{Performance comparison between the baseline strategy and our FPMM strategy.} \label{baseline}
\resizebox{86mm}{10mm}{
  \begin{tabular}{|@{}c@{}|@{\,}c@{\,}|@{}c@{}|}
  \hline
  & Baseline Strategy & Our FPMM Strategy  \\ \hline
Recovery Threshold & $P'\!\!=\!\!3K\!+\!\max\{T_{\mathbf{A}}\!+\!T_{\mathbf{B}}\}\!-\!2$ &  $P\!=\!4K+T_{\mathbf{A}}+T_{\mathbf{B}}-3$  \\ \hline
  Download Cost & $\lambda\gamma P'$ &  $\lambda\gamma P$  \\ \hline
 Server Comp. & ${O}(RV\lambda\gamma)$ & ${O}(\frac{R\lambda\omega+V\omega\gamma}{K}+\frac{\lambda\omega\gamma}{K})$ \\ \hline
  Decoding Comp. & ${{O}}(\lambda\gamma P'(\log P')^2\log\log P')$ & $O(\lambda\gamma P(\log P)^2\log\log P)$  \\ \hline
Storage Cost & $RV\lambda\gamma$ & $\frac{R\lambda\omega+V\omega\gamma}{K}$  \\ \hline
  \end{tabular}}
   \begin{tablenotes}
       \footnotesize
       \item[] Here we set $L=M=1$ to minimize  the recovery threshold of our FPMM strategy.
\end{tablenotes}
\end{table}

\subsection{Comparison for PSMM and FPMM with Colluding Constraints}\label{subsection:collcom}
The state-of-the-art strategies for PSMM and FPMM problems with colluding constraints are constructed in \cite{PSMM_Zhu}. The PSMM and FPMM problems considered in this paper reduce to the classical PSMM and T-FPMM problems with colluding constraints respectively, when data is simply replicated across $N$ servers (i.e., setting $K=1$), and the same privacy constraints are enforced (i.e., setting $S=T$ for our PSMM or $T_{\mathbf{A}}=T_{\mathbf{B}}$ for our FPMM).  In this special case, the complexities of the proposed PSMM and FPMM strategies are consistent with the ones in \cite{PSMM_Zhu}, when the matrices associated to $\mathbf{A}$ are only divided horizontally and the matrices associated to $\mathbf{B}$ are only divided vertically.

The work in \cite{PSMM_Zhu} presents a general transformation  from secure matrix multiplication (SMM) to private matrix multiplication, and the strategies in \cite{PSMM_Zhu} are constructed following the structure inspired by the encoding functions of secure matrix multiplication (SMM) strategies \cite{Rouayheb_secure_code,EP_SMC,Zhu_SDMM,yang2018secure,PSMM:2}. 
Specifically, one SMM strategy for computing the product of matrices $\mathbf{A}'$ and $\mathbf{B}'$ includes the encoding functions of $\mathbf{A}'$ and $\mathbf{B}'$. For the PSMM problem with replication storage, the strategy in \cite{PSMM_Zhu} directly exploits the encoding function of $\mathbf{A}'$ to create secret shares of the confidential matrix $\mathbf{A}$, and exploits the encoding function of $\mathbf{B}'$ to create private queries for each of matrices in $\mathbf{B}^{(1)},\ldots,\mathbf{B}^{(V)}$, such that the response computed at each server resembles the response computed in the SMM strategy. The core point is that the matrices $\mathbf{B}^{(1)},\ldots,\mathbf{B}^{(V)}$ are stored at each server in a replicated form, so that the queries for the matrices $\mathbf{B}^{(1)},\ldots,\mathbf{B}^{(V)}$ can be constructed by reusing the encoding function of the matrix $\mathbf{B}'$ in SMM strategy and then each server is able to generate an arbitrary desired response  using its local replication storage. 
When MDS coded storage is considered, it is highly challenging to generalize the idea in \cite{PSMM_Zhu} to enable that, 
as the server just stores encoded data of the matrices $\mathbf{B}^{(1)},\ldots,\mathbf{B}^{(V)}$ at this time. See Remark \ref{MDS:effect} for the effect of MDS-coded storage on the design of PSMM strategies. The FPMM strategy follows similar arguments. 

\begin{figure}[htbp]
	\centering
		\subfigure[PSMM Strategy]
		{ \label{figure:privacy:PSMM}
		\includegraphics[width=0.225\textwidth]{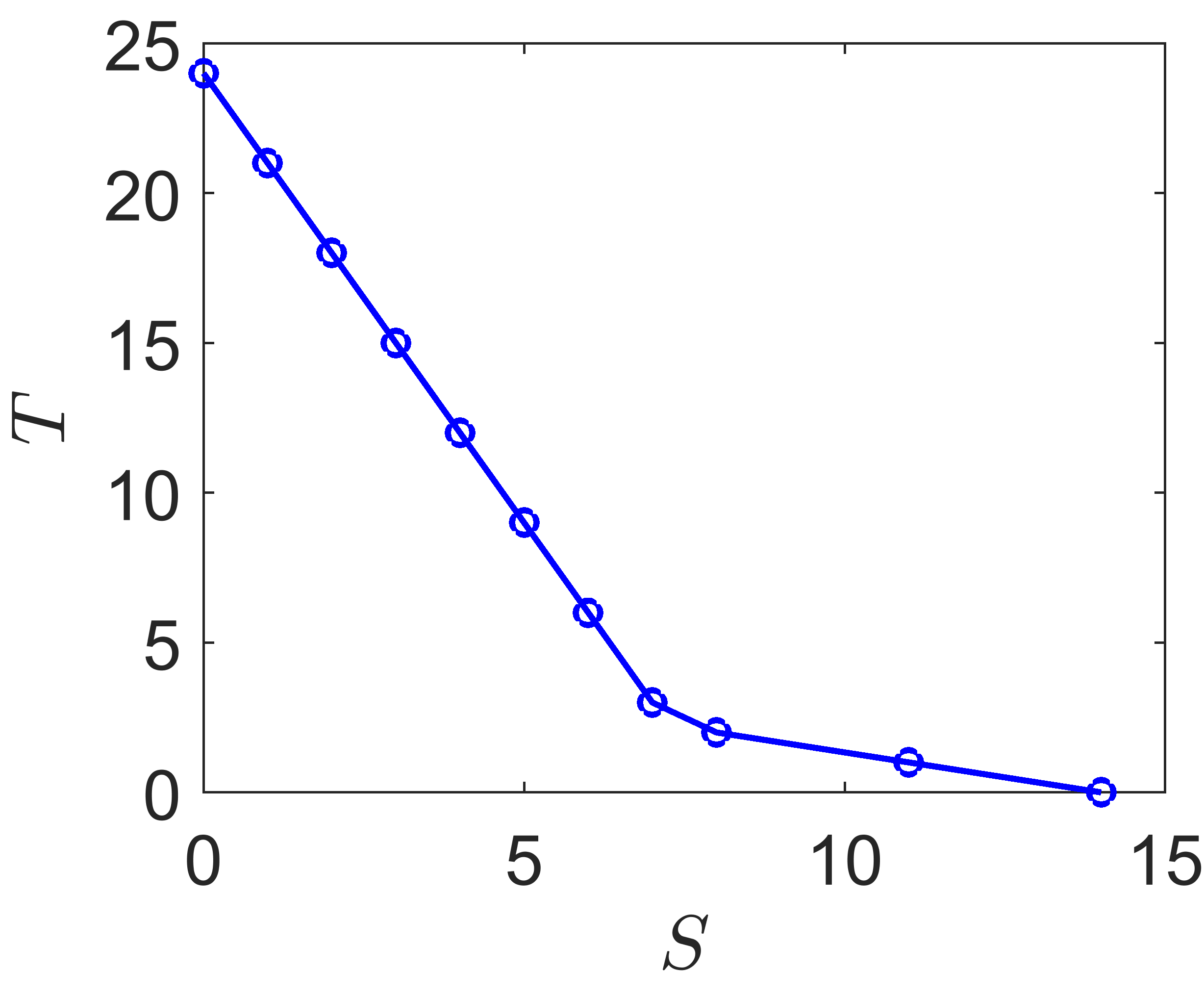}
		}
		\subfigure[FPMM Strategy]
		{ \label{figure:privacy:FPMM}
		\includegraphics[width=0.225\textwidth]{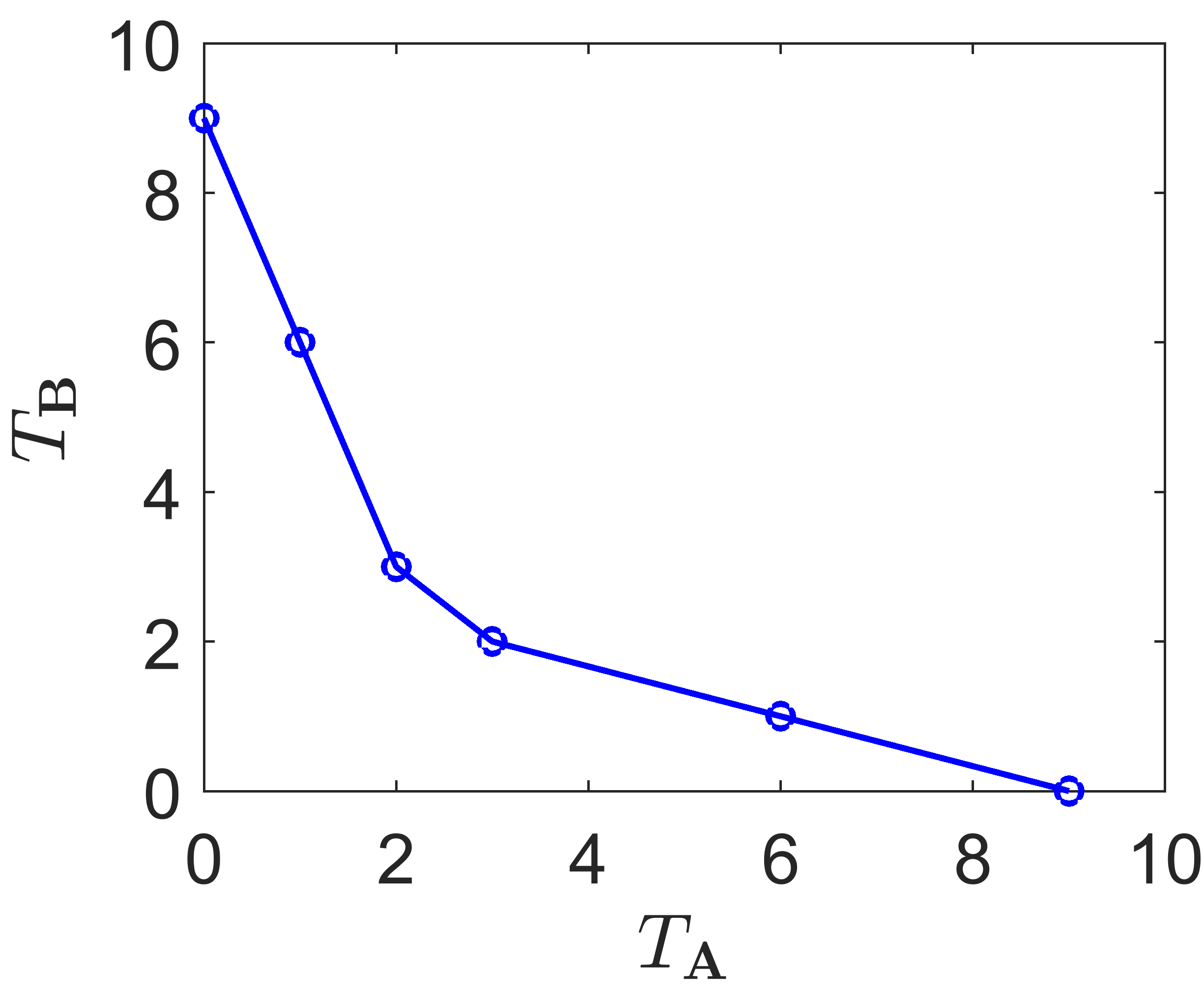}
		}
	\caption{(a) Achievable secrecy parameter $S$ for matrix $\mathbf{A}$ and privacy parameter $T$ for the matrix index $\theta$, for the proposed PSMM strategy, and (b) Achievable privacy parameter $T_{\mathbf{A}}$ for matrix index $\theta_{\mathbf{A}}$ and privacy parameter $T_{\mathbf{B}}$ for another matrix index $\theta_{\mathbf{B}}$, for the proposed FPMM strategy, with $K=6,L=M=3$ and $P=100$.}
	\label{figure:privacy}
\end{figure}

Furthermore, we can observe from Theorem \ref{theorem:PSMM} and \ref{theorem:FPMM} that,  given any recovery threshold $P$, there is a tradeoff between the secrecy parameter $S$ for matrix $\mathbf{A}$ and the privacy parameter $T$ for the matrix index $\theta$ in the proposed PSMM strategy; and a tradeoff between the privacy parameter $T_{\mathbf{A}}$ for the matrix index $\theta_{\mathbf{A}}$ and the privacy parameter $T_{\mathbf{B}}$ for another matrix index $\theta_{\mathbf{B}}$ in the proposed FPMM strategy. We numerically illustrate this tradeoff in Fig. \ref{figure:privacy}.


\section{Conclusion}\label{conclusion}
In this paper, we design novel strategies for the problems of PSMM and FPMM, with information-theoretic secrecy and privacy guarantees.
The key idea is to construct secret shares of a confidential matrix and private indices, from polynomials whose degree parameters are carefully selected to satisfy the intended computations. The considered settings of PSMM and FPMM include all previously private matrix multiplication problems as special cases. When applying the proposed PSMM strategy to solve the PSMM problem with weaker privacy guarantees, we obtain major advantages in reducing the overall communication and computation complexities. 




\bibliographystyle{ieeetr}
\bibliography{reference.bib}

\end{document}